\documentclass[a4paper,twocolumn]{article}
\pdfoutput=1

\usepackage[english]{babel}
\usepackage[utf8x]{inputenc}
\usepackage[T1]{fontenc}
\usepackage{balance}
\usepackage{float}
\usepackage{abstract}
\usepackage{amsfonts}
\usepackage{caption}
\usepackage{tikz}
\usepackage[a4paper,top=1.5cm,bottom=1.5cm,left=1.5cm,right=1.5cm,marginparwidth=1.75cm]{geometry}
\usepackage{amsmath}
\usepackage{graphicx}
\usepackage[colorinlistoftodos]{todonotes}
\usepackage[colorlinks=true, allcolors=blue]{hyperref}
\usepackage{authblk}
\usepackage[bottom]{footmisc}
\usepackage{widetext}
\usepackage{latexsym}
\usepackage{wasysym}
\usepackage{amssymb}
\usepackage{float}
\usepackage{multicol}

\title{Quasi-normal modes of black holes and naked singularities: revisiting the WKB method}
\author[1,4,a]{Edison C. Santos}
\author[2,3,b]{J\'{u}lio C. Fabris}
\author[4,c]{Jos\'{e} A. de Freitas Pacheco}

 \affil[1]{{\small PPGCosmo, CCE, Universidade Federal do Esp\'irito Santo, 29075-910, Vit\'{o}ria, ES, Brasil}}
 \affil[2]{{\small N\'{u}cleo Cosmo-ufes, Departamento de F\'{i}sca, CCE, Universidade Federal do Esp\'irito Santo, 29075-910, Vit\'{o}ria, ES, Brasil}}
 \affil[3]{{\small National Research Nuclear University MEPhI, 115409, Moscow, Russia}}
\affil[4]{{\small Observatoire de la C\^{o}te d'Azur, BP 4229, 06034 Nice Cedex 4, France}}

\affil[a]{{\small edison\_cesar@hotmail.com}}
\affil[b]{{\small julio.fabris@cosmo-ufes.org}}
\affil[c]{{\small pacheco@obs-nice.fr}}

\begin{document}

\twocolumn[
  \begin{@twocolumnfalse}
    \maketitle
    \begin{abstract}
\noindent In this paper we revisit the analysis of the \textit{ringdown} frequencies in the form of quasi-normal modes for Schwarzschild, Schwarzschild de-Sitter and Reissner-Nordstr{\"o}m space-times. We plot these frequencies, using the third-order WKB semi-analytical method against the mass, charge and cosmological constant for each corresponding space-time for various different spin-fields. The results indicate the stability of each black hole solution, including the extremal Reissner-Nordstr{\"o}m space-time.
Finally, we discover four different traits for the Schwarzschild de-Sitter space-time: i) the frequencies vanish in the extremal case; ii) the frequencies swap behavior before and after the extremal value; iii) indication that the naked singularity is stable for positive mass; iv) for the naked singularity case, there exists a cut-off mass in the scalar-field perturbation which depends only on the multipole number $\ell$.

\vspace{0.5cm}
    \end{abstract}
  \end{@twocolumnfalse}
  ]
  
\maketitle

\section{Introduction}


It is well established that general relativity is a theory plagued by singularities. When these singularities are within  \textit{event horizons} \cite{Hawkingbook}, we call the resulting structure a black hole.

In agreement with the \textit{no-hair theorem} \cite{PhysRev.164.1776}, the classical black hole solutions may be obtained with dependence solely on its mass $M$, charge $Q$ and the angular momentum parameter $a$, aside from the cosmological constant, $\Lambda$, that may be directly introduced into the field equations. These solutions can be classified as a sub-class of the \textit{Kerr-Newman (anti) de-Sitter} space-time.

Any of the given solutions may be perturbed as
\begin{equation} \label{General perturbation}
	g_{\mu \nu}'(x^\lambda) = g_{\mu \nu}(x^\lambda) + h_{\mu \nu}(x^\lambda),
\end{equation}
where $g_{\mu \nu}(x^\lambda)$ is the black hole background metric and $h_{\mu \nu}(x^\lambda)$ is the perturbed metric. To guarantee the validity of the linearized field equations, the following condition must be satisfied 
\begin{equation}
	\bigg\lvert \frac{h_{\mu \nu}(x^\lambda)}{g_{\mu \nu}(x^\lambda)} \bigg\rvert \ll 1,
\end{equation}
that is, the perturbation term is never dominant over the background metric.

These perturbations can be used to describe the last moment of a system of binary objects, such as two black holes, in the process of coalescence. This process is roughly divided into three different stages, each with their own unique features: the first stage is the \textit{inspiral phase}, where both bodies are spiraling inward due to the emission of gravitational waves, albeit far enough from each other to be distinguished as two objects. The horizons then coincide and the \textit{merger phase} begins; this is where the strongest gravitational waves are emitted. The modeling of this phase requires the computation of the full-solution of Einstein's non-linear equations, which is only possible through numerical analysis. The remnant of this collision is a single vibrating black hole, with its frequency decaying exponentially. The last stage of the coalescence is known as the \textit{ringdown phase} and is well described by perturbation theory.

Perturbation theory was first applied to black hole physics with \cite{Regge1957} and was picked up more than a decade later by \cite{Zerilli1970}. The main result obtained in these papers was the reduction of the cumbersome set of perturbed equations into a single Schr\"{o}dinger-like equation for the radial perturbations. Subsequently, the first paper to effectively calculate quasi-normal modes, that is, the damped oscillation signal of a black hole, was \cite{Vishveshwara1970}. Thereafter, the calculation of quasi-normal modes in a wide variety of contexts and methods was initiated (\cite{Konoplya2011}, \cite{Berti2009}, \cite{Kokkotas1999} and references therein).

In this work we reproduce some results with multiple goals: first to thoroughly analyse how the frequencies are affected by the parameters of the black holes; secondly, when within the limitation of the method, to further extend the analysis for naked singularities, in the case of Schwarzschild de-Sitter case, and extremal black holes, for the charged Reissner-Nordstr{\"o}m black hole, as these cases are not always discussed in the literature. In this investigation we re-obtained the non-emission of Hawking radiation for the Schwarzschild de-Sitter solution, and also the following original results: the frequencies swap their behavior before and after reaching its extremal limit; an indication that the naked singularity case is stable for a positive mass; the existence of a cut-off value for $\Lambda$ and mass for the naked singularity, in a scalar field perturbation.

In what follows, we review the basics of black hole perturbation theory in section \ref{Section 2}; in section \ref{Section 3} we describe the first-order WKB method to solve the master equation for the radial perturbations, and then generalize it up to third-order; in section \ref{Section 4} we present the results obtained for each solution, and examine them separately, giving detailed attention to Schwarzchild de-Sitter, where four different characteristics are discussed in depth, as we explore the new results of this work; finally, in section \ref{Summary}, the main results are summarized.
\section{Black hole perturbation theory}
\label{Section 2}

Black hole perturbation theory began with the pioneering work of \cite{Regge1957}, where they investigated whether Schwarzschild's metric was stable under the general non-spherical perturbation, as described in equation \eqref{General perturbation}. 

Any second-order symmetric tensor may be decomposed into its Scalar, Vector and Tensor parts (known as \textit{SVT decomposition}). These quantities may be assembled into two orthogonal classes: either \textit{polar} or \textit{axial}\footnote{Some authors may call it \textit{even} and \textit{odd} parity, respectively. For a table with other conventions used throughout the literature, see appendix B in \cite{Zerilli1970}. }; where the former is invariant under rotation and the latter is not. The explicit difference between both modes is that under parity transformations, the spherical harmonic index $\ell$ transforms as $(-1)^\ell$ for the \textit{polar} class and $(-1)^{\ell +1}$ for the \textit{axial} class. In the linear-order, both modes decouple and a simpler set of equations is acquired.

Due to the spherical symmetry and staticity, the most general metric $g_{\mu\nu}$ can always be arranged in the form\footnote{This is true in the context of general relativity. Such format may not be the most general in alternative theories of gravity.}
\begin{equation} \label{ds2 depending on f(r)}
	ds^2 = f(r)dt^2 - f(r)^{-1} dr^2 - r^2 d\Omega^2,
\end{equation}
where the function $f(r)$, which is also known as the \textit{lapse function}, is characterized differently for each space-time. 

The main result obtained in \cite{Regge1957} shows that a variable decomposition of the type
\begin{equation} \label{Variable decomposition}
	\Psi (t, r, \theta, \phi) = e^{i \omega t} \psi(r) Y(\theta, \phi)
\end{equation}
yields a Schr\"{o}dinger-like equation for the function $\psi(r)$ given by
\begin{equation} \label{master equation}
	\frac{d^2 \psi(r)}{d r_*^2} + (\omega^2 - V(r)) \psi(r) = 0.
\end{equation}
Where $r_*$ is defined as the tortoise coordinate $\frac{dr}{dr_*} = f(r)$ and $Y(\theta, \phi)$ as the spherical harmonics functions. The effective potential, $V(r)$, depends on the characteristics of the given black hole solution and on the type of the perturbation, which can be either \textit{axial} or \textit{polar}. As we will see in the next section, the radial-perturbation equation of every classical black hole solution, with the exception of the rotating ones,\footnote{For Kerr and Kerr-Newman solutions, the method must be changed. The radial equation obtained is in the form of a Teukolsky equation, first introduced in \cite{Teukolsky:1973ha}.} can be expressed in this specific format.

\subsection{Metrics and effective potentials}
The study of black holes, through perturbation theory, turns out to be a problem of analysing the effective potential in equation \eqref{master equation}. In this section, we review the line elements of equation \eqref{ds2 depending on f(r)}, with their respective effective potentials, for various different types of perturbations.

\subsubsection{Schwarzschild}
\label{Subsection Schwarzschild}
The simplest black hole solution is described by the Schwarzschild space-time. The \textit{lapse function} $f(r)$ is
\begin{equation} \label{f Schwarzschild}
	f_{Sch}(r) = 1 - \frac{2M}{r},
\end{equation}
with $M$ being its mass. For a positive mass, a black hole solution is obtained within the Schwarzschild radius, $r_{Sch} = 2 M$. If the mass turns out to be negative, the space-time describes a \textit{naked singularity} solution. Thence the interpretation of $M$ would not be that of the mass, which recovers the Newtonian limit at spatial infinity, instead, we refer to it as the \textit{mass parameter}. It is widely accepted that the latter should not exist in nature, this is known as the \textit{cosmic censorship conjecture} \cite{Penrose:1969pc}.

As explained previously, both types of perturbation obey equation \eqref{master equation}, however, the effective potentials are different. In \cite{Regge1957}, we can see that the effective potential for \textit{axial} perturbations is
\begin{align} \label{Potential axial Schwarzschild}
	V^a_{Sch}(r) = f_{Sch} \left[ \frac{\ell(\ell+1)}{r^2} + 2 \frac{\beta M}{r^3} \right],
\end{align}
which is known as the \textit{Regge-Wheeler potential}. The parameter $\ell$ is a constant related to the angular momentum and $\beta$ to the spin of the field through $\beta = 1-S^2$, also called the \textit{field's spin weight} \cite{Berti2009}.

Despite the perturbation due to different spins being characterized by the same equation\footnote{At least for the \textit{axial} perturbation of Schwarzschild solution.}, they originate from very different natures. For $S=2$, we get \textit{gravitational perturbations}, which arise from the perturbation of the Einstein tensor itself. The \textit{electromagnetic perturbations}, i.e. $S=1$, comes from the analysis of the electromagnetic four-potential $A_\mu$ when subjected to Maxwell's equation in a given space-time \cite{Fernando:2017qrd}. The four-potential may also be decomposed into its \textit{axial} and \textit{polar} parts, using  the same method employed by Regge and Wheeler.

Finally, for $S=0$, we have \textit{scalar perturbation} that results from the behavior of the massless Klein-Gordon equation
\begin{equation}
	\Box \phi(r) = 0,
\end{equation}
in the curved space-time. For any (electro-) vacuum space-time described by equation \eqref{ds2 depending on f(r)}, the effective potential  is
\begin{equation}\label{Scalar field easy general potential}
	V^s(r) = f(r) \left[ \frac{\ell (\ell +1)}{r^2} + \frac{f'(r)}{r} \right],
\end{equation}
where the prime denotes a derivative in respect to the radial coordinate $r$. For the Schwarzschild \textit{lapse function}, the result obtained is the one described in equation \eqref{Potential axial Schwarzschild} for $S=0$ (i.e. $\beta=1$).

On the other hand, the effective potential for the \textit{polar} perturbations, known as \textit{Zerilli potential} \cite{Zerilli1970}, is given as
\begin{equation}
  \begin{aligned} \label{Potential polar Schwarzschild}
        V^p_{Sch}(r) = \frac{f_{Sch}}{r^3 (\eta r + 3M)^2} [2\eta^2 (\eta+1)r^3 + \\
         6\eta^2 M r^2 + 18 \eta M^2 r + 18 M^3],
  \end{aligned}
\end{equation}
with $2\eta = (\ell-1)(\ell+2)$.

Furthermore, in \cite{Chandrasekhar1984} we find that the two potentials for gravitational perturbations, described by equations \eqref{Potential axial Schwarzschild} and \eqref{Potential polar Schwarzschild}, are related by 
\begin{equation} \label{Superpotentials}
	V^{\overset{a}{p}}_{Sch} (r) = \pm \alpha \frac{dW}{dr_*} + \alpha^2 W^2 + \kappa W
\end{equation}
with
\begin{equation}
	\begin{aligned}
		W = f_{Sch} \frac{1}{2r^3 (\ell r + 3M)}, && \beta = 6M, && \kappa = 4\eta(\eta+1).
	\end{aligned}
\end{equation}
This relation between potentials is called the \textit{super-partner potentials}, which is a typical term in the context of supersymmetric theories \cite{Cardoso:2001bb}. Such property arises from the fact that these potentials possess the same amplitudes for both the transmitted and reflected waves \cite{Chandrasekhar_book}. This property is known as the \textit{isospectral relation}, since they have the same quasi-normal spectrum. Potentials with this structure imply a relation between the wave-functions $\psi^a$ and $\psi^p$ of the type
\begin{equation} \label{Superwavefunctions}
	\psi^{\overset{a}{p}} = \frac{1}{\alpha - \omega^2} \left( \mp W + \frac{d}{dr_*} \right) \psi^{\overset{p}{a}},
\end{equation}
hence, after obtaining either the frequencies or the wave-function for one potential, the other can be easily derived \cite{Berti2009}.

\subsubsection{Reissner-Nordstr\"{o}m}
The line element describing a charged black hole with charge $Q$ is known as the Reissner-Nordstr\"{o}m space-time. The event horizons are situated at $r_{\pm} = M \pm \sqrt{M^2-Q^2}$, and thus three different cases are possible: i) $M > Q$ gives a black hole where the outer horizon is an event horizon and the inner one a Cauchy horizon; ii) $M = Q$ has both horizons coinciding and is called the \textit{extremal solution}; iii) $Q > M$ admits no horizons and a charged \textit{naked singularity} space-time is described. 

This space-time follows the structure stated in equation \eqref{ds2 depending on f(r)}, with
\begin{equation}
	f_{RN}(r) = 1 - \frac{2M}{r} + \frac{Q^2}{r^2} .
\end{equation}
The \textit{polar} perturbation potential is given by
\begin{align} \label{RN potential polar}
	V_{j(RN)}^p = \frac{f_{RN}}{r^3} \left[ \ell(\ell+1)r - q_i + \frac{4 Q^2}{r} \right],
\end{align}
where $i,j = 1, 2 \quad  (i \neq j)$ and with the following definitions
\begin{equation}
	\begin{aligned} \label{q_i}
        q_1 &= 3 M  +  \sqrt{9 M^2 + 4(\ell-1)(\ell+1)Q^2}, \\
        q_2 &= 3 M  -  \sqrt{9 M^2 + 4(\ell-1)(\ell+1)Q^2}.
	\end{aligned}
\end{equation}

Further, the \textit{axial} perturbation effective potential is
\begin{equation} \label{RN potential axial}
	V^a_{j(RN)} = \frac{f_{RN}}{r^3} \left[U \pm \frac{1}{2}(q_1-q_2) J\right],
\end{equation}
with
\begin{equation}
	\begin{aligned}
    \bar{\omega} &= \eta r + 3M - 2\frac{Q^2}{r}, \\
    J &= f_{RN} \frac{r}{\bar{\omega}}^2 (2\eta r + 3M) + \frac{\eta r + M}{\bar{\omega}}, \\
	U &= (2\eta r + 3M) J + (\bar{\omega} - \eta r - M) - \frac{2\eta r^2}{\bar{\omega}} f_{RN}.
	\end{aligned}
\end{equation}
As for the Schwarzschild solution, both \textit{axial} and \textit{polar} effective potentials are correlated via
\begin{equation} \label{RN potential relation}
	V_j^{p} = V_j^{a} + 2q_i \frac{d}{dr_*} \left[ \frac{f_{RN}}{r [(\ell-1)(\ell+2)r + q_i]} \right].
\end{equation}

Using relation \eqref{Scalar field easy general potential}, the effective potential for a \textit{scalar perturbation} is given by
\begin{equation}
    V_{RN}^s (r) = f_{RN} \left[ \frac{\ell(\ell +1)}{r^2} + 2\frac{M}{r^2} - 2 \frac{Q^2}{r^3} \right]
\end{equation}
We see that if we choose $Q=0$, we re-obtain the \textit{scalar perturbation} for the Schwarzschild effective potential, defined in \eqref{Potential axial Schwarzschild} for $S=0$ (i.e. $\beta = 1$).

From equations \eqref{RN potential polar} and \eqref{RN potential axial}, it is clear that differently from the uncharged solution, each perturbation type has two different potentials, instead of a single one. Since the perturbation of Einstein's and Maxwell's equations cannot be fully disentangled, that is, there is no purely electromagnetic nor gravitational modes of oscillation, there will be an emission of both electromagnetic and gravitational radiation for all kinds of perturbation \cite{Kokkotas1988}.

\subsubsection{Schwarzschild de-Sitter}
\label{Sds explanation}
Black holes may also be investigated when immersed within a universe with a cosmological constant, $\Lambda$. This solution bears the name of its discoverer: Kotler, and may have either a positive or negative value for $\Lambda$ \cite{Kottler1918}. Despite it being Kotler's discovery, the solution is more commonly known as \textit{Schwarzschild de-Sitter} (positive $\Lambda$) or \textit{anti de-Sitter} (negative $\Lambda$).

This space-time is described by
\begin{align}
	f_{SdS}(r) &= 1 - \frac{2M}{r} - \frac{\Lambda r^2}{3}.
\end{align}
Since there are two different parameters, $M$ and $\Lambda$, there is a diversity of solutions, as summarized in table \ref{TableSdS}.

\begin{table}[ht]
\resizebox{\columnwidth}{!}{%
\begin{tabular}{|c|c|c|c|}
\hline
                               & \textbf{0 horizon} & \textbf{1 horizon} & \textbf{2 horizons} \\ \hline
\textbf{$M>0$ and $\Lambda>0$} & $9\Lambda M^2 > 1$  & $1 = 9\Lambda M^2$   & $0 < 9\Lambda M^2 < 1$    \\ \hline
\textbf{$M>0$ and $\Lambda<0$} & Never              & Always             & Never               \\ \hline
\textbf{$M<0$ and $\Lambda>0$} & Never              & Always             & Never               \\ \hline
\textbf{$M<0$ and $\Lambda<0$} & Always             & Never              & Never               \\ \hline
\end{tabular}
}
\caption{The different number of horizons are displayed for different values of the cosmological constant, $\Lambda$, and mass, $M$, for the Schwarzschild (anti) de-Sitter solution.}
\label{TableSdS}
\end{table}

In the case of Schwarzschild anti-de Sitter, if the mass parameter is positive, only one horizon solution is possible, which is an event horizon. Then, this space-time describes an observer outside of a black hole in an contracting universe. This solution is a crucial element for the AdS/CFT\footnote{Anti de Sitter/Conformal Field Theory.} correspondence \cite{Maldacena:1997re}. However, if the mass parameter is negative we are left with a horizonless case, which is a \textit{naked singularity} space-time.

A much richer set of solutions is possible for a positive cosmological constant. The horizons may be localized via the following relations, \cite{Faraoni:2018xwo}
\begin{equation}
    \begin{aligned}
            r_1 &= \frac{2}{\sqrt{\Lambda}} \sin(\psi), \\
            r_2 &= \sqrt{\frac{3}{\Lambda}} \cos(\psi) - \frac{1}{\sqrt{\Lambda}} \sin(\psi), \\
            r_3 &= \sqrt{\frac{3}{\Lambda}} \cos(\psi) - \frac{1}{\sqrt{\Lambda}} \sin(\psi).
    \end{aligned}
\end{equation}
with
\begin{equation}
    \sin(3\psi) = 3 \sqrt{\Lambda} M.
\end{equation}
However, for a negative mass parameter only a cosmological horizon is present. Then, the observer would be found between the naked singularity and the cosmological horizon. If the mass parameter is positive, three different possibilities are possible: i) in the range $0 < 9 \Lambda M^2<1$ both cosmological and event horizons are present, hence an observer is bounded by them both, and is found between a black hole and an expanding universe; ii) when $9 \Lambda M^2=1$, both horizons coincide and an \textit{extremal} black hole is created, also known as the \textit{Nairiai limit} \cite{1951SRToh..35...46H}; iii) lastly, if $9 \Lambda M^2 > 1$, no horizons are formed, and this can be interpreted as if the event horizon has grown larger than the cosmological one, making it effectively disappear, and hence the singularity is not screened \cite{Faraoni:2018xwo}.

Perturbation theory was first applied to the Schwarzschild de-Sitter solution in \cite{PhysRevD.42.2577}, but we refer the reader to \cite{Zhidenko2003}, where the whole set of effective potentials is summarized. For this case, the effective potential of the \textit{axial} perturbation is
\begin{align} \label{Effective SdS axial}
	V^a_{SdS}(r) &= f_{SdS} \left[ \frac{\ell(\ell+1)}{r^2} + 2 \frac{\beta M}{r^3} \right], && S=1,2
\end{align}
where, once again, $\beta = 1 - S^2$. The gravitational perturbation of the \textit{polar} type is then
\begin{equation} \label{SdS polar potential}
	\begin{aligned}
        V^p_{SdS}(r) &= \frac{f_{S\Lambda}}{r^3 (\eta r + 3M)^2} [2\eta^2 (\eta+1)r^3 + \\
         & 6\eta^2 M r^2 + 3M^2(3\eta r - \Lambda r^3) + 18 M^3]. 
	\end{aligned}
\end{equation}

The same structure encountered in Schwarzschild relating both potentials, i.e. equations \eqref{Superpotentials} and \eqref{Superwavefunctions}, also holds true for Schwarzschild de-Sitter. The adjusted definitions of the function $W$ and the constant $\beta$ can be found in \cite{Zhidenko2003}.

The perturbation for a scalar-field, $S=0$, may be obtained using equation \eqref{Scalar field easy general potential} and is given by
\begin{equation}
	V^s_{SdS}(r) = f_{SdS} \left[ \frac{\ell(\ell+1)}{r^2} + 2 \frac{ M}{r^3} - \frac{2}{3} \Lambda \right].
\end{equation}
It is clear from the previous equations that when setting $\Lambda = 0$ the results for Schwarzschild are re-obtained.








\section{The WKB method}
\label{Section 3}

The WKB method (named after the physicists Gregor Wentzel, Hendrik Kramers and L\'{e}on Brillouin) is an approximate method to solve linear differential equations. The most important and recognizable usage of this method is to solve the time-independent Schr\"{o}dinger equation \cite{Landau_QM}
\begin{equation} \label{Schrodinger equation}
	- \frac{\hbar^2}{2M} \frac{d^2 \Psi(x)}{dx^2}  + (V(x) - E) \Psi(x) =0.
\end{equation}
Due to the resemblance of equation \eqref{Schrodinger equation} with the master equation \eqref{master equation}, the idea to employ the same method for black hole perturbation theory is evident.

Formally, the WKB method consists of an approximation in a single exponential power series of the type
\begin{equation}
	\psi(r) \approx \exp \left[ \frac{1}{\epsilon} \sum_{n=0}^{\infty} \epsilon^n S_n(r) \right], \qquad \epsilon \to 0,
\end{equation}
and the differential equation to be analysed has the following general form
\begin{equation} \label{General equation for Q}
	\frac{d^2 \psi(r)}{dr^2} = Q(r) \psi(r), \qquad Q(r) \neq 0.
\end{equation}
The main difference between a quantum mechanical problem, in which you may have different numbers of returning points for the effective potential (although it is usually a single one), is that in the black hole perturbation theory the function $Q(r)$ contains two turning points, necessarily. For this reason, the matching procedure must be altered, as was developed by \cite{WKBbook}.

\vspace{0.5cm}
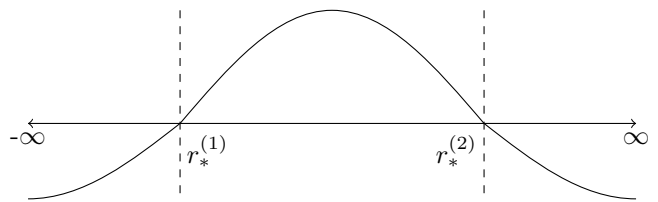
\begin{figure}[ht]
\begin{tikzpicture}
  \draw [<->] (-4,0) node[below] {-$\infty$} -- (-2,0) node[below right,circle] {$r_{*}^{(1)}$}-- (2,0) node[below left,circle] {$r_{*}^{(2)}$} -- (4,0) node[below] {$\infty$};
  \draw[dashed] (-2,1.5) -- (-2,-1);
  \draw[dashed] (2,1.5) -- (2,-1);
  \draw (-2,0) sin (0,1.5) cos (2,0);
  \draw (-4,-1) cos (-2,0);
  \draw (2,0) sin (4,-1);
\end{tikzpicture}
	\caption{A sketch of a general potential $V(r)$ with two turning points, $r_*^{(1)}$ and $r_*^{(2)}$.}
    \label{WKB potential sketch}
\end{figure}

The main peculiarity in a quasi-normal mode problem is that specific boundary conditions must be set. Since waves cannot emerge from the Minkowskian spatial infinity, a purely outgoing waves boundary condition must be chosen
\begin{equation} \label{BC1}
	\psi(r) = Z_{out}\psi(r)_{out}, \qquad r_* \to \infty,
\end{equation}
while at the event horizon, since waves cannot arise from within the black hole, only incoming waves are permitted, then
\begin{equation} \label{BC2}
	\psi(r) = Z_{in} \psi(r)_{in}, \qquad r_* \to - \infty.
\end{equation}
With these boundary conditions it is clear that the black hole itself is losing energy in the form of gravitational waves. This happens either by waves going into the event horizon or by dissipating themselves in the spatial infinity.

Using these appropriate boundary conditions, the WKB matching condition throughout the three regions leads to
\begin{equation}
	\frac{Q_0}{\sqrt{2Q_0''}} = i (n+1/2), \qquad n \in \mathbb{N},
\end{equation}
where the subscript $0$ represents the function $Q(r)$ evaluated at its maximum and will be omitted from now on in order to simplify the notation. The factor $n$ is known as the \textit{overtone number}, which is a discrete quantity, and the fundamental oscillation mode is given by  $n=0$.

Until this point, the method is fully general for any differential equation that obeys equation \eqref{General equation for Q}, for any given two turning point function $Q(r)$, with $Q''(r) \neq 0$, where the latter implies that a maximum must exist.


For black holes physics, the method was first used by \cite{1983mgm..conf..599M}, where he used $Q(r) = V(r) - \omega^2$, and found the real and imaginary frequencies of the oscillations to be described as
\begin{equation} \label{WKB condition first order}
	\omega^2 = V - i (n+1/2) \sqrt{-2 V^{(2)}},
\end{equation}
with the derivatives of the potentials given by
\begin{equation}
	V^{(m)} = \frac{d^m V}{dr_*^m} = f(r) \frac{d}{dr} V^{m - 1},
\end{equation}
which must be evaluated in the potential's maximum. 

The easiness of the WKB method is well stated in equation \eqref{WKB condition first order}: instead of solving a rather difficult differential equation numerically, the frequencies are obtained by solving a much simpler one. Furthermore, it can easily be expanded to further orders of approximation by directly adding more terms to the condition above.

This method has been first expanded to the third-order in \cite{Iyer1987a}. Subsequently, a succession of papers arose studying every classical black hole solution: Schwarzschild \cite{Iyer1987}, Reissner-Nordstr\"{o}m \cite{Kokkotas1988} and Kerr \cite{Seidel1990}. Thereafter, it has been even further developed to the sixth-order for an $N-$dimensional Schwarzschild black hole \cite{Konoplya2003}.

In the third-order approximation, the condition given in equation \eqref{WKB condition first order} changes in the following manner
\begin{equation}\label{frequencies third order}
	\omega^2 =  [V + \sqrt{(-2 V^{(2)})} \Gamma] - i \sqrt{\alpha (-2 V^{(2)})} (1+ \Omega)
\end{equation}
whereas the cumbersome quantities $\Gamma$ and $\Omega$ are
\begin{equation}
\begin{aligned}
		\Gamma &= \frac{1}{\sqrt{-2V^{(2)}}} \Biggr[ \frac{1}{8} \left(\frac{V^{(4)}}{V^{(2)}} \right) \left( \frac{1}{4} +\alpha \right) \\
		& \hspace{3.2cm}-\frac{1}{288} \left( \frac{V^{(3)}}{V^{(2)}} \right)^2 (7 + 60\alpha) \Biggr],
\end{aligned}
\end{equation}        
\begin{equation}
\begin{aligned}
        \Omega &= -\frac{1}{2 V^{(2)}} \bigg[ \frac{5}{6912} \left( \frac{V^{(3)}}{V^{(2)}} \right)^4 (77 + 188 \alpha)  \\ 
        & \qquad - \frac{1}{384} \left( \frac{(V^{(3)})^2 V^{(4)}}{(V^{(2)})^3} (51 + 100 \alpha) \right)  \\ 
        & \qquad + \frac{1}{2304} \left( \frac{V^{(4)}}{V^{(2)}} \right)^2 (67+68\alpha) \\
        & \qquad + \frac{1}{288} \left( \frac{V^{(3)} V^{(5)}}{(V^{(2)})^2} \right) (19+28\alpha)  \\
        & \qquad - \frac{1}{288} \left( \frac{V^{(6)}}{V^{(2)}} \right) (5+4\alpha) \bigg]
\end{aligned}
\end{equation}
with $\alpha = (n+1/2)^2$.

In table \ref{TableComparison}, we compare the WKB method that we have described and implemented, with the numerical method applied in \cite{Chandrasekhar1975b} and to Leaver's semi-analytic method \cite{Leaver1985}. The results were acquired using a Schwarschild black hole, with unity mass, $M=1$, for the \textit{polar} perturbation in units of $2 M \omega$.

\begin{table*}[ht!]
\resizebox{\textwidth}{!}{
\begin{tabular}{cc|ccc|ccc}
\hline
Multipole      & Overtone  & Chandrasekhar & Leaver & 3rd-order WKB & \multicolumn{3}{c}{Error (\%)}                                \\
  number                                     &               number                   &                                &                         &                                & Leaver-Chandra & Chandra-3rd & Leaver-3rd  \\
\hline                                       
$\ell$=2 & n=0                         & 0.74734 -0.17792i                & 0.747343 -0.177925i            & 0.746324 -0.1784348i    & 0.000 -0.003i                  & 0.136 -0.289i       & 0.137 -0.286i   \\
                                       & n=1                              & 0.69687 -0.54938i              & 0.693422 -0.54783i      & 0.692034 -0.54983i             & 0.497 -0.283i       & 0.699 -0.082i    & 0.201 -0.364i   \\
                                       & n=2                              &                                & 0.602107 -0.956554i     & 0.60587 -0.942128i             &                     &                  & 0.621 -1.531i   \\
                                       & n=3                              &                                & 0.50301 -1.410296i      & 0.494924 -1.345796i            &                     &                  & 1.634 -4.793i   \\
\hline
$\ell$=3 & n=0                         & 1.19889 -0.18541i                & 1.198887 -0.185406i            & 1.19853 -0.1854568i     & 0.000 -0.002i                  & 0.030 -0.025i       & 0.030 -0.027i   \\
                                       & n=1                              & 1.16402 -0.56231i              & 1.165288  -0.562596i    & 1.16471 -0.562812i             & 0.109 -0.051i       & 0.059 -0.089i    & 0.050 -0.038i   \\
                                       & n=2                              & 0.85257 -0.74546i              & 1.10337 -0.958186i      & 1.1064 -0.953368i              & 22.730 -22.201i     & 22.942 -21.808i  & 0.274 -0.505i   \\
                                       & n=3                              &                                & 1.023924 -1.380674i     & 1.031494 -1.354858i            &                     &                  & 0.734 -1.905i   \\
\hline
$\ell$=4 & n=0                         & 1.61835 -0.18832i                & 1.61836 -0.18832i              & 1.618196 -0.1883422i    & 0.001 -0.000i                  & 0.010 -0.012i       & 0.010 -0.012i   \\
                                       & n=1                              & 1.59313 -0.56877i              & 1.59326 -0.56886i       & 1.592998 -0.568732i            & 0.008 -0.016i       & 0.008 -0.007i    & 0.016 -0.023i   \\
                                       & n=2                              & 1.12019 -0.84658i              & 1.54542 -0.95982i       & 1.547272 -0.957948i            & 27.515 -11.798i     & 27.602 -11.626i  & 0.120 -0.195i   \\
                                       & n=3                              &                                & 1.47968 -1.36784i       & 1.486624 -1.3566i              &                     &                  & 0.467 -0.829i   \\ \cline{2-8} 
\hline
\end{tabular}
}
\caption{We compare the numerical method used by Chandrasekhar \cite{Chandrasekhar1975b}, Leaver's semi-analytic method \cite{Leaver1985} (for $\ell=4$ values, see \cite{Nollert1999}) and our implementation of the third-order WKB method. The results were acquired using the Schwarschild black hole, with unity mass, $M=  1$, for the \textit{polar} perturbation in units of $2M\omega$.}
\label{TableComparison}
\end{table*}

\subsection{Availability of the method for naked singularities}

Naked singularities arise in the space-times analyzed here in the following forms: for Schwarzschild when the mass parameter is negative; for Reissner-Nordstr\"{o}m when either the mass parameter is negative or the inequality $Q > M$ is satisfied; lastly, for Schwarschild (anti) de-Sitter, as stated in table \ref{TableSdS}, it occurs for $M < 0$ and $\Lambda < 0$, or if $ 9 \Lambda M^2 > 0$, assuming that both parameters are positive.

\begin{figure}[H]
    \includegraphics[width=\textwidth/2]{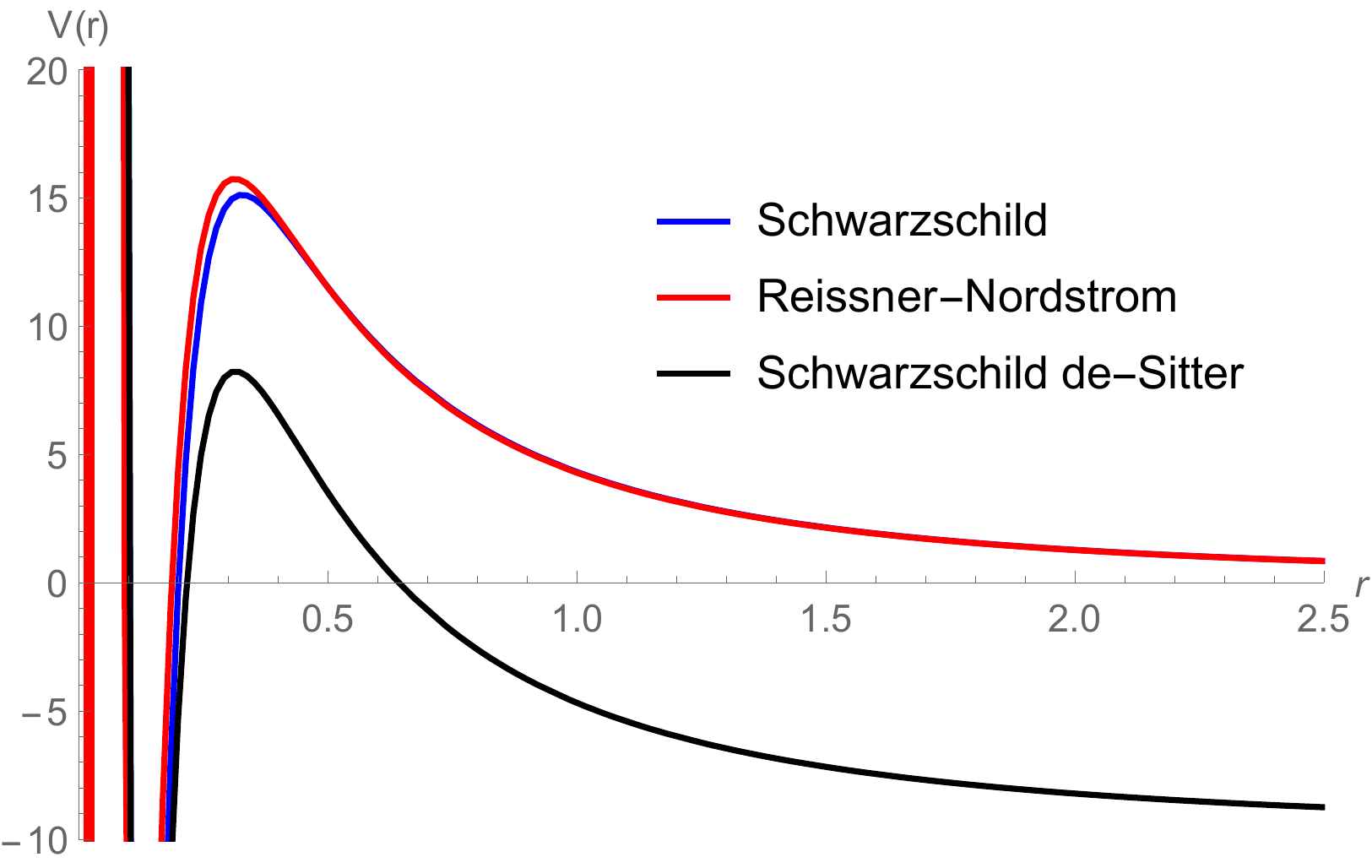}
  \caption{Plot of the effective potentials of the black hole solutions of the Schwarzschild, Reissner-Nordstr\"om and Schwarzschild de-Sitter space-time, for \textit{axial}, \textit{polar} and \textit{axial gravitational perturbations}, respectively. It is clear that every potential possess two turning points and a maximum, as necessary for the WKB method to be applied. The parameters used were: $M = 0.1$, $Q = 0.05$, $\ell = 2$, $\Lambda = 5$. }
  \label{Plots: Pot1}
\end{figure}

For the WKB method, as described here, to be properly applied, the following conditions must be fulfilled:
\begin{enumerate}
    \item The potential must possess two turning points;
    \item The potential must have a maximum;
    \item Purely in-going waves arriving to the center and outgoing waves departing in the spatial infinity.
\end{enumerate}
It is clear from figure \ref{Plots: Pot1}, that the black hole solutions of the space-times analyzed here satisfies every single condition needed. Hence, the WKB method may be correctly applied to these cases.

\begin{figure}[H]
    \includegraphics[width=\textwidth/2]{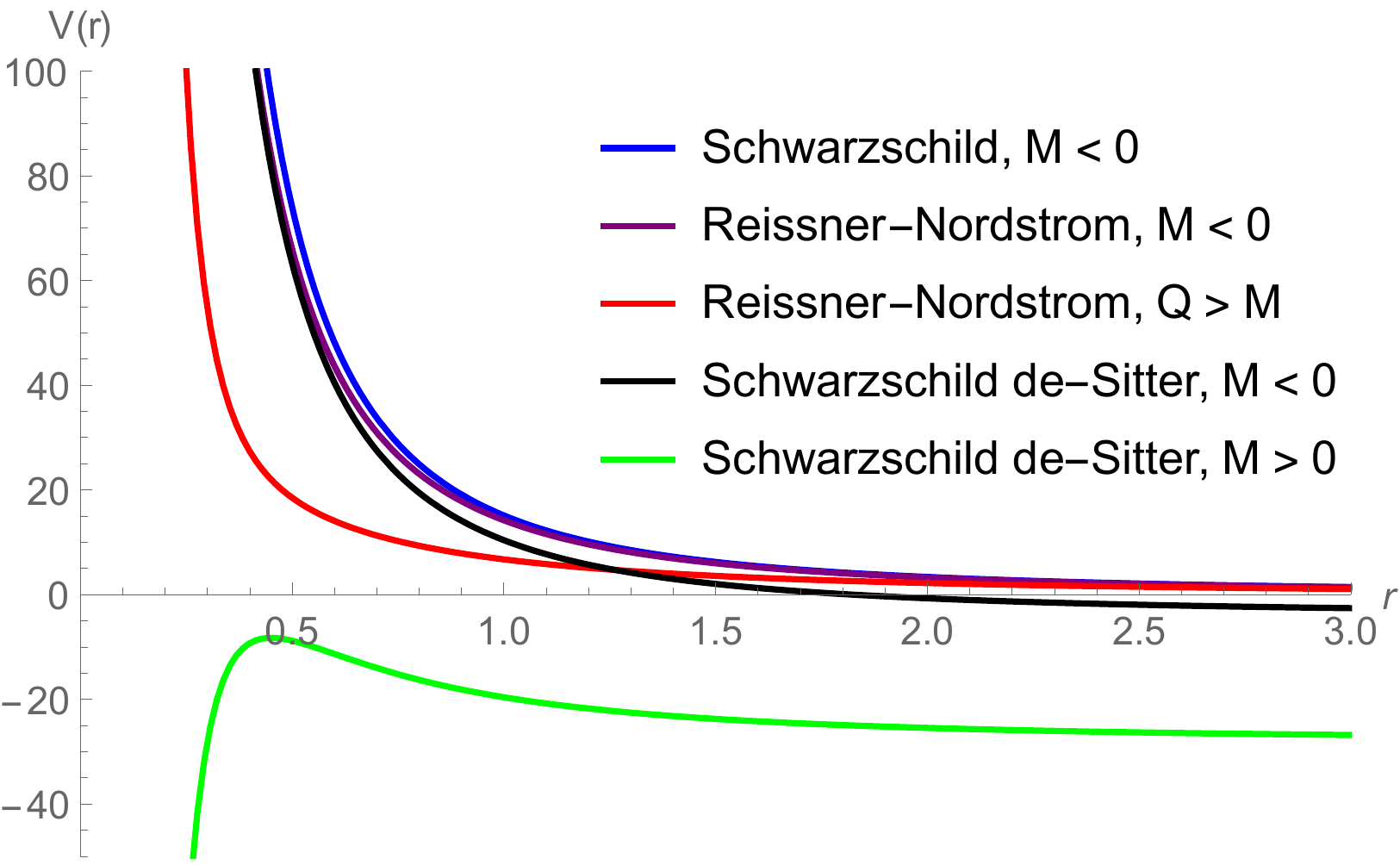}
  \caption{Plot of the effective potentials of the naked singularities solutions of the Schwarzschild , Reissner-Nordstr\"om and Schwarzschild de-Sitter space-time, for \textit{axial gravitational}, \textit{polar gravitational} and \textit{axial electromagnetic perturbations}, respectively. It is clear that neither of these cases possess a extrema nor two turning points, besides for the Schwarzschild de-Sitter naked singularity with positive mass space-time, represented by the green curve; due to this, the WKB method may be applied only for the latter case. The parameters used were $\ell = 3$ and: for Schwarzschild $M = -0.1$; for Reissner-Nordstr\"om with negative mass $M = 0.1$ and $Q = 0.07$, and for the case in which $Q > M$, we set $M = 0.2$ and $Q = 0.3$; for Schwarzschild de-Sitter with negative mass $M = -0.1$ and $\Lambda = 1$ and for positive mass $M = 0.15$ ad $\Lambda = 7$, which satisfies the naked singularity case stated in $9 \Lambda M^2 > 1$.}
  \label{Plots: Pot2}
\end{figure}

However, every naked singularity case cited in the previous paragraph fails in each item, as can be seen in figure \ref{Plots: Pot2}: they do not contain two turning points, nor a maximum. Lastly, since the potential diverges to $+\infty$ as $r \to 0$, it does not allow in-going waves into the naked singularity. 



However that is not the case for Schwarzschild de-Sitter naked singularity with positive mass. In figure \ref{Plots: Pot2}, by direct inspection it is clear that the green curve resembles the effective potentials displayed in figure \ref{Plots: Pot1}, which represents the typical effective potential behavior. This case fulfill every condition for the WKB method to be utilized: it possesses a maximum and it has two turning points. The part that requires detailed attention is the third condition, as stated in equations \eqref{BC1} and \eqref{BC2}. At first, it seems that since the tortoise coordinate for the naked singularity case does not cover the whole range of the reals, the boundary condition in equation \eqref{BC1} would not be satisfied and the method could not be applied. 

Even with all of this, we must make it clear that not every effective potential for Schwarzschild de-Sitter naked singularity with positive mass satisfy all these conditions. The best example is the \textit{axial gravitational} perturbation in equation \eqref{Effective SdS axial}, which clearly diverges to $+ \infty$ as the radial coordinate approaches zero. 

But, following the deduction from \cite{Schutz}, the requirements for the matching condition, equation \eqref{WKB condition first order}, to be satisfied are given solely by $r_* \ll r_*^{(1)}$ and $r_* \gg r_*^{(2)}$, where both $r_*^{(1)}$ and $r_*^{(2)}$ are the turning points\footnote{To be precise, they perform an expansion of the solution of the parabolic cylinder differential equation, but in the asymptotic form for large $|r_* |$.}. That is, there is no need for the tortoise coordinate to cover the whole range $-\infty < r_* < \infty $, as long as the turning points are far from the extremes of the potential. Strictly speaking, the condition of purely in-going waves is still satisfied, which characterizes an attractive potential.

With this we conclude that the method can be correctly applied for positive mass Schwarzschild de-Sitter naked singularity space-time. For the solutions considered in this paper, this is the only naked singularity case where the WKB method may be correctly used.

\subsection{Stability condition}
\label{Subsection Stability}

The variable decomposition in equation \eqref{Variable decomposition}, i.e. the wave function $\Psi(t, r, \theta, \phi)$, has a time functional form of the type
\begin{equation}
	\Psi(t, r, \theta, \phi) \approx e^{i \omega t},
\end{equation}
which describes an oscillation in time. However, when analysing equation \eqref{frequencies third order}, we see that the general frequency also has an imaginary component
\begin{equation}
	\Psi(t, r, \theta, \phi) \approx e^{(i \omega_r + \omega_i)t}.
\end{equation}
In the above equation, the real and imaginary part of $\omega$ has been decomposed using a different sign convention $\omega = \omega_r - i \omega_i$. 

The real part of the frequency represents the real oscillation of the black hole, which is positive definite. Conversely, the imaginary part of the frequency may be either positive or negative. If $\omega_i > 0$, the exponential term diverges as time evolves, and it follows that the oscillation of the black hole will always grow, thus describing an \textit{unstable solution}. On the other hand, for $\omega_i < 0$, we have a damped oscillation, since the whole exponential term decreases to zero as the time coordinate infinitely increases and, therefore, the black hole ceases its oscillation; this describes a \textit{stable solution}. Finally, $\omega_i = 0 $ describes a normal mode, with a black hole infinitely vibrating.

Hence, we can define a quasi-normal mode as a general oscillation which possess an exponential damping factor. Note that this is only true when $\omega_i < 0$.

\section{Results}
\label{Section 4}

In this section, we discuss the results we obtained using the WKB method up to the third-order of the first few \textit{overtone} numbers, $n$, and multipole values, $\ell$. We find that most of the plots follow the same general behaviour and, as such, we only present a fraction of them. We believe that the plots displayed in this paper summarize well the main results that we obtained.\footnote{Specific cases may be requested to the authors via e-mail.}

In figures \ref{Plots: SaScalar}, \ref{Plots: RNV1aGrav}, \ref{Plots: SdSaScalar} and \ref{Plots: SdSaEM} the graphs on the left side are plotted for $n=0$ and $\ell = [1,2,3]$, while the ones on the right are for $n = [0,1,2]$ and $\ell=1$. In the case of the \textit{polar} gravitational perturbations, depicted in figures \ref{Plots: SpGrav}, \ref{Plots: RNV2pGrav} and \ref{Plots: SdSpGrav}, the intervals for the parameters in the graphs on the left are $n=0$ and $\ell = [2,3,4]$, and for the ones on the right are $n = [0,1,2]$ and $\ell=2$. The continuous lines represent the real frequencies and the dashed ones are the respective imaginary frequencies. When not explicitly stated in the figures, we adopted the following values for the constants: $M=1$, $Q=0.5$ and $\Lambda = 10^{-1}$. 

\subsection{Schwarzschild}
\begin{figure*}[!ht]
  \begin{tabular}{cc}
  \includegraphics[width=\textwidth/2]{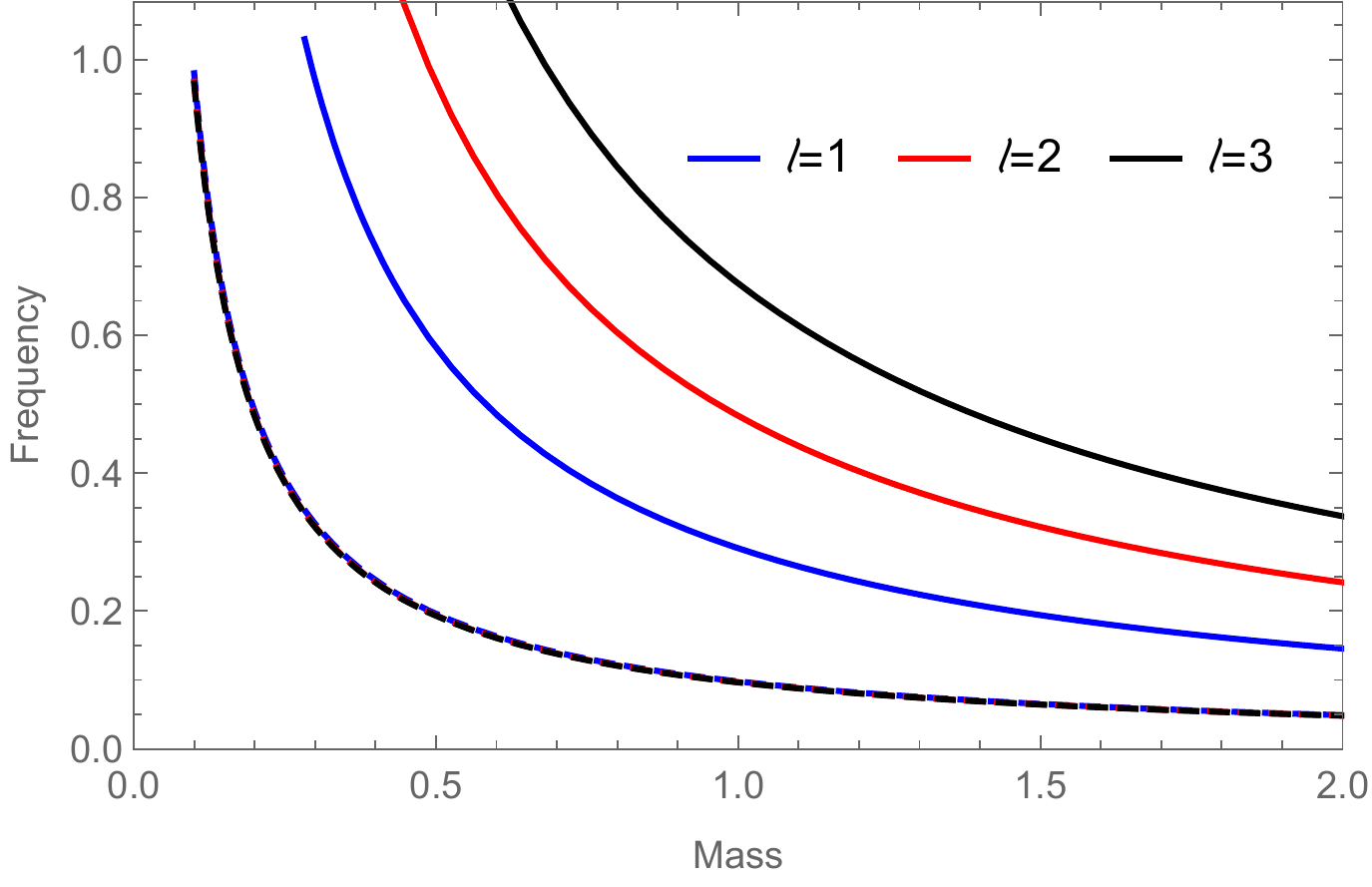} &  \includegraphics[width=\textwidth/2]{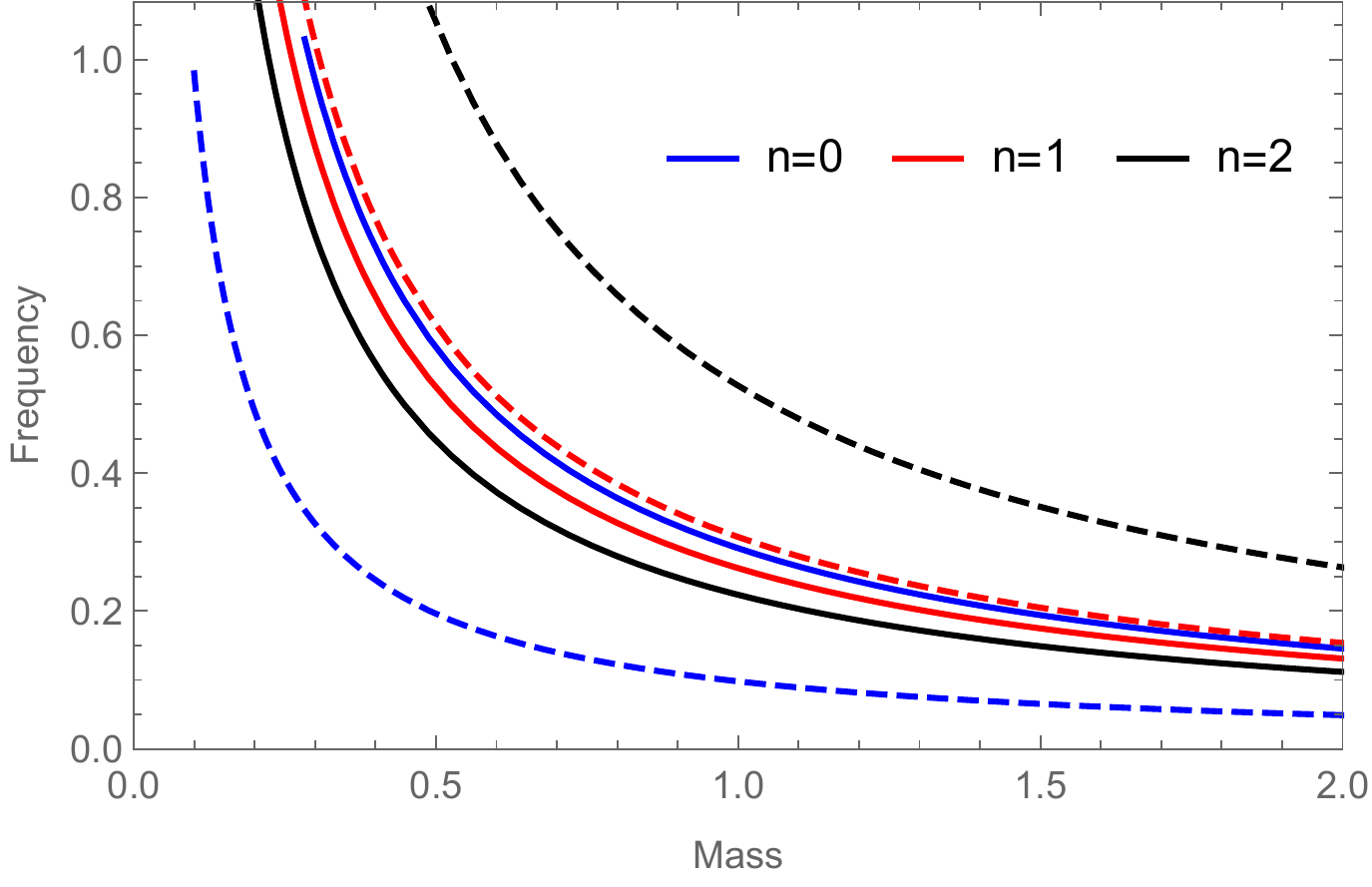}
  \end{tabular}
  \caption{
  \textit{Axial} perturbation of a $S=0$ field in the Schwarzschild space-time for a varying mass. The continuous lines represent the real frequencies and the dashed ones are the respective imaginary frequencies. The plot on the left maintains $n=0$ constant while $\ell$ varies through $1$ to $3$. We see that as $\ell$ increases the real frequencies also grow, however the imaginary frequencies seems insensible to such change. The plot on the right maintains $\ell = 1$ constant while $n$ varies through $0$ to $2$. We can identify that the real frequencies decrease as $n$ increases, while the opposite occurs to the imaginary frequencies. }
  \label{Plots: SaScalar}
\end{figure*}

\begin{figure*}[!ht]
  \begin{tabular}{cc}
  \includegraphics[width=\textwidth/2]{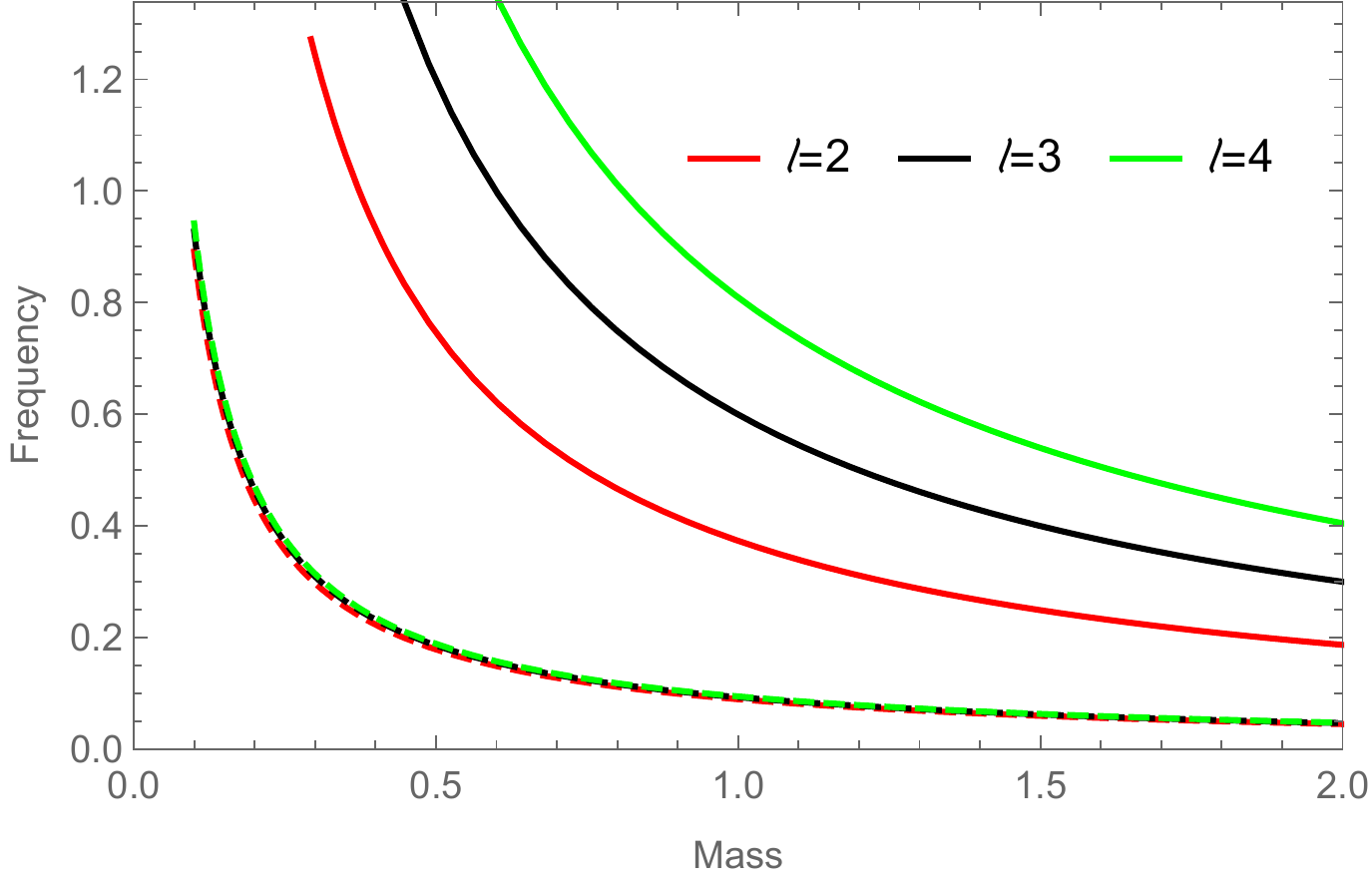} &  \includegraphics[width=\textwidth/2]{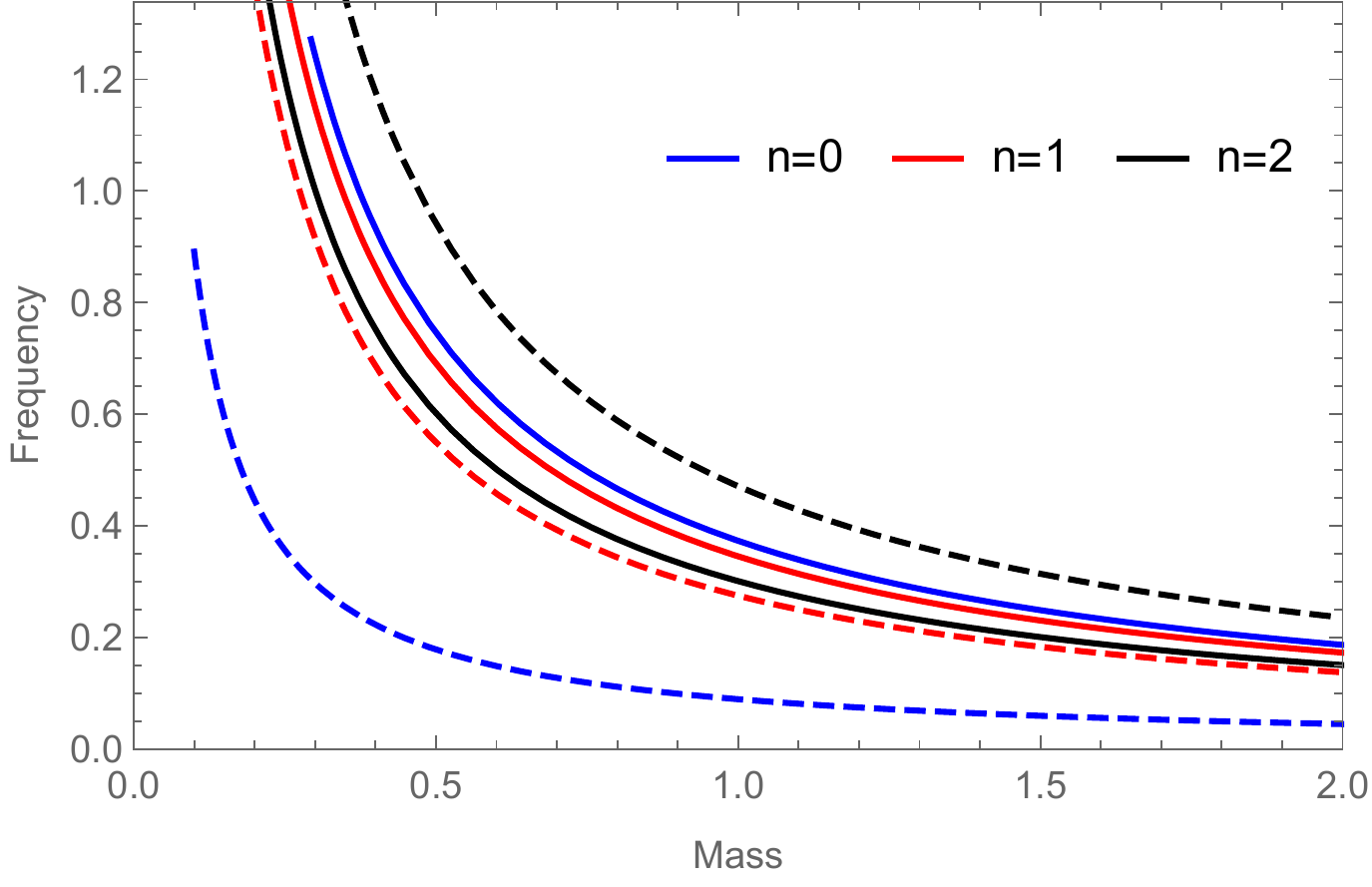}
  \end{tabular}
  \caption{\textit{Polar} perturbation of a $S=2$ field in Schwarzschild space-time for a varying mass. The continuous lines represent the real frequencies and the dashed ones are the respective imaginary frequencies. The plot on the left maintains $n=0$ constant while $\ell$ varies through $2$ to $4$. We see that as $\ell$ increases the real frequencies also grow, however the imaginary frequencies seem insensible to such changes. The plot on the right maintains $\ell = 2$ constant while $n$ varies through $0$ to $2$. We can identify that the real frequencies decrease as $n$ increases, while the opposite occurs to the imaginary frequencies.  }
    \label{Plots: SpGrav}
\end{figure*}

The plots in figure \ref{Plots: SaScalar} and figure \ref{Plots: SpGrav} are the results that we obtained for the perturbation of the Schwarzschild solution, using the formalism explained in subsection \ref{Subsection Schwarzschild}. The former shows the \textit{axial} perturbation of a scalar field and the latter is for the \textit{polar} gravitational perturbation. We can see that for a fixed \textit{overtone} number, the imaginary frequencies remain almost unchanged. On the other hand, as it increases, and $\ell$ remains fixed, both real and imaginary frequencies are altered. As the value of $n$ grows the frequency decreases, as does the damping factor. It is also clear that, as the mass of the black hole increases, its vibration is reduced and the damping effect becomes less pronounced.

In the last three columns of table \ref{TableComparison}, it is plain that the errors of the two semi-analytic methods become larger as the pair $\ell$ and $n$ increases, in comparison with the numerical integration. However, we find that the error is less than $0,3\%$ for the fundamental modes. When comparing the semi-analytic methods in the last column, it is evident that the error of the WKB method worsens as the value of $n$ increases, with a value of $4.79\%$ for the imaginary part given $\ell = 2$ and $n=3$. As $\ell$ grows, our results get increasingly more accurate, with $0.01 -0.012i \%$ for $\ell = 4$. Hence, our results are compatible with the expected behaviour of WKB method per se \cite{Nikolaos2005, Berti2009}.

Finally, we can conclude from both figures \ref{Plots: SaScalar} and \ref{Plots: SpGrav} as well as from table \ref{TableComparison}, that the stability condition as defined in \ref{Subsection Stability} is satisfied and, therefore, the black hole is stable. Every result that we have obtained for the Schwarzschild solution corroborates with what is already known in the literature.

The case of the naked singularity in the Schwarzschild space-time cannot be analysed via the WKB method. None of the effective potentials satisfies the condition stated in Figure \ref{WKB potential sketch}, i.e. there are no two turning points, hence the potential has no maximum and the whole method should be modified to adjust for such condition.

\subsection{Reissner-Nordstr{\"o}m}

In what follows, we discuss the results obtained for the Reissner-Nordstr{\"o}m space-time.

\begin{figure*}[!ht]
  \begin{tabular}{cc}
  \includegraphics[width=\textwidth/2]{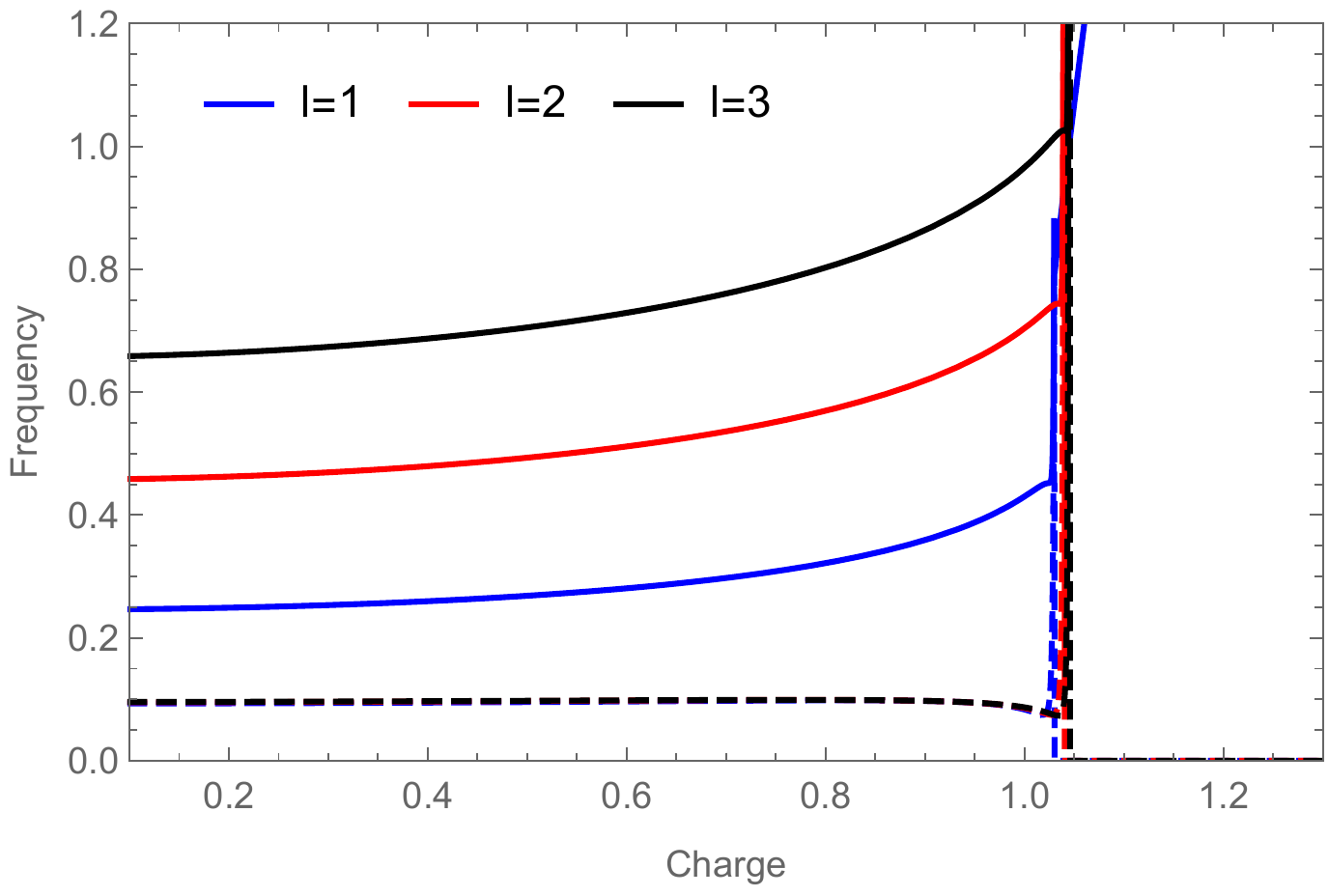} &  \includegraphics[width=\textwidth/2]{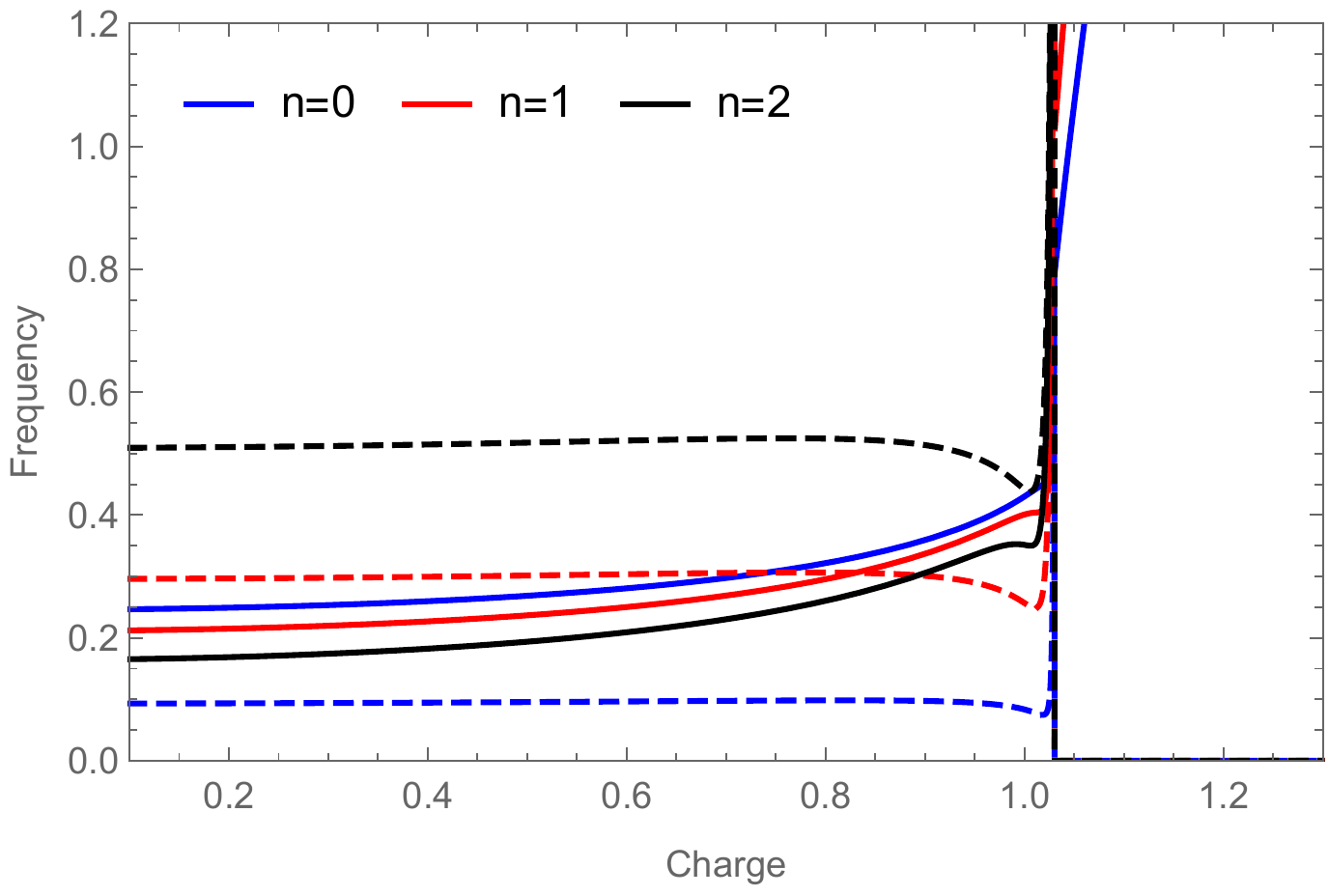}
  \end{tabular}
  \caption{\textit{Axial} perturbation of potential $V_1$ of a $S=2$ field in Reissner-Norstr\"{o}m space-time for a fixed mass $M=1$.
  The continuous lines represent the real frequencies and the dashed ones are the respective imaginary frequencies. The plot on the left maintains $n=0$ constant while $\ell$ varies through $1$ to $3$. We see that as $\ell$ increases the real frequencies also grow, however the imaginary frequencies seems insensible to such change, exactly as in the Schwarzschild case. The plot on the right maintains $\ell = 1$ constant while $n$ varies through $0$ to $2$. We can identify that the real frequencies lower their value as $n$ increases, while the opposite occurs to the imaginary frequencies. In both cases we note that the imaginary frequencies are mostly insensible to the variation of the electric charge of the black hole, at the same time, it increases the real oscillation frequency.}
  \label{Plots: RNV1aGrav}
\end{figure*}

In figure \ref{Plots: RNV1aGrav}, we have an \textit{axial} perturbation of the potential $V_1$, where we allow the charge to vary until after its extremal value, for a fixed mass. We can see that the frequencies abruptly diverge when the inequality $Q>M$ becomes satisfied, which supports the \textit{cosmic censorship conjecture}. We were also able to check that the imaginary frequencies are weakly dependent on the black hole charge, as was first identified in \cite{Cardoso:2017soq}. Lastly, we see that the real frequencies of the black hole increase as the charges grows, that is, its charge does not act as a damping factor, as the mass does, but as a driving parameter. 

\begin{figure*}[!ht]
  \begin{tabular}{cc}
  \includegraphics[width=\textwidth/2]{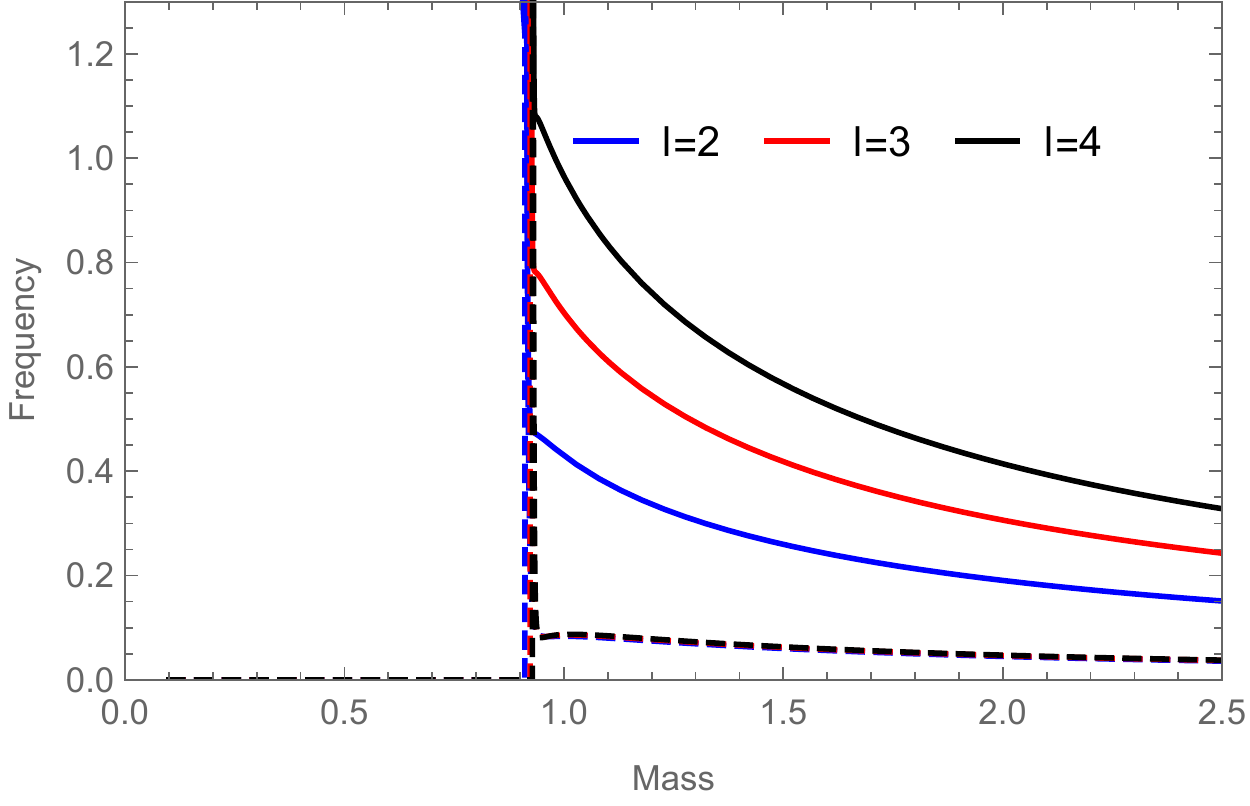} &  \includegraphics[width=\textwidth/2]{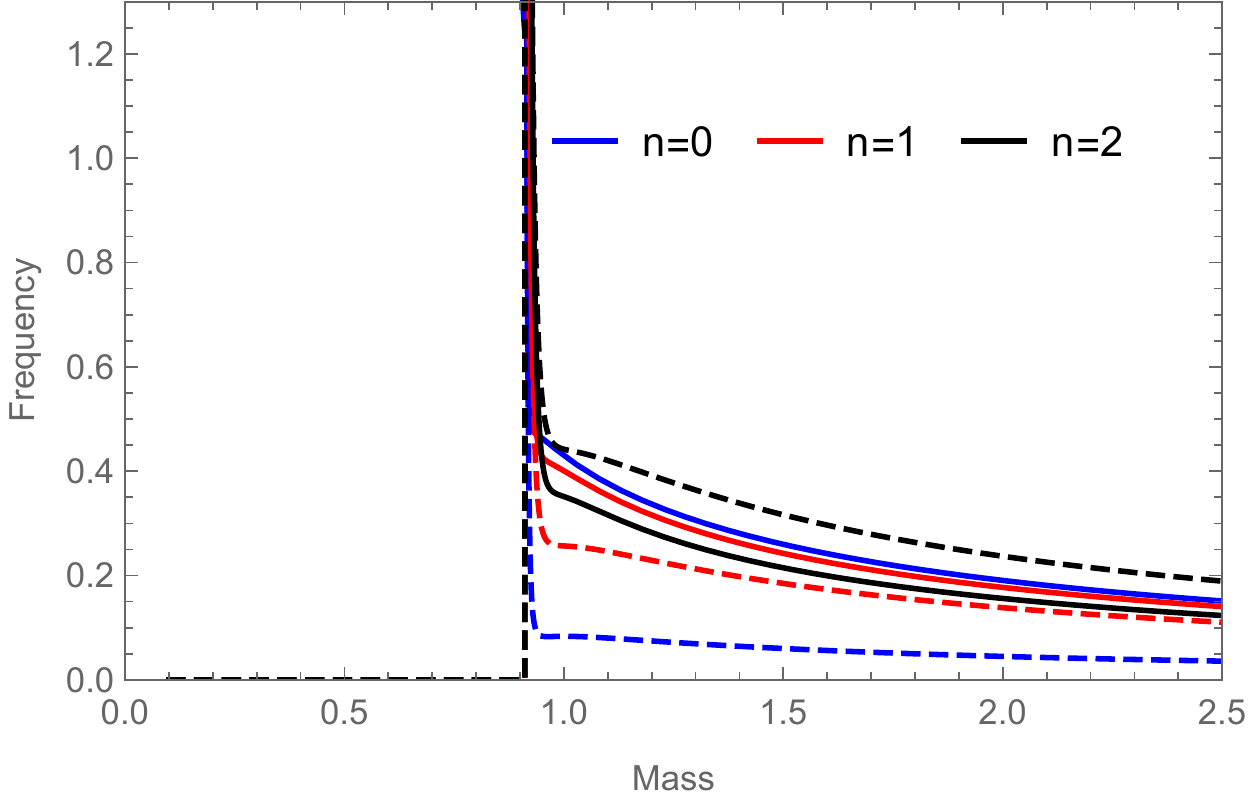}
  \end{tabular}
  \caption{\textit{Polar} perturbation of potential $V_2$ of a $S=2$ field in Reissner-Norstr\"{o}m space-time for a fixed charge $Q=1$.
  The continuous lines represent the real frequencies and the dashed ones are the respective imaginary frequencies. The plot on the left  maintains $n=0$ constant while $\ell$ varies through $2$ to $4$. We see that as $\ell$ increases the real frequencies also grow, however the imaginary frequencies seems insensible to such change, exactly as in the Schwarzschild case. The plot on the right maintains $\ell = 2$ constant while $n$ varies through $0$ to $2$. We can identify that the real frequencies lower their value as $n$ increases, while the opposite occurs to the imaginary frequencies.}
  \label{Plots: RNV2pGrav}
\end{figure*}

In figure \ref{Plots: RNV2pGrav}, we display a \textit{polar} perturbation for the $V_2$ potential, for a fixed and for a variable charge parameter. Note that the non-existence of frequencies on the leftmost part of the plots, $Q>M$, are, once again, due to the \textit{cosmic censorship conjecture}. After the point in which $Q=M$, the frequencies begin to closely resemble the Schwarzschild solution, with the mass damping the frequencies and the same analysis described in the previous subsection applies.

\begin{figure*}[!ht]
  \begin{tabular}{cc}
  \includegraphics[width=\textwidth/2]{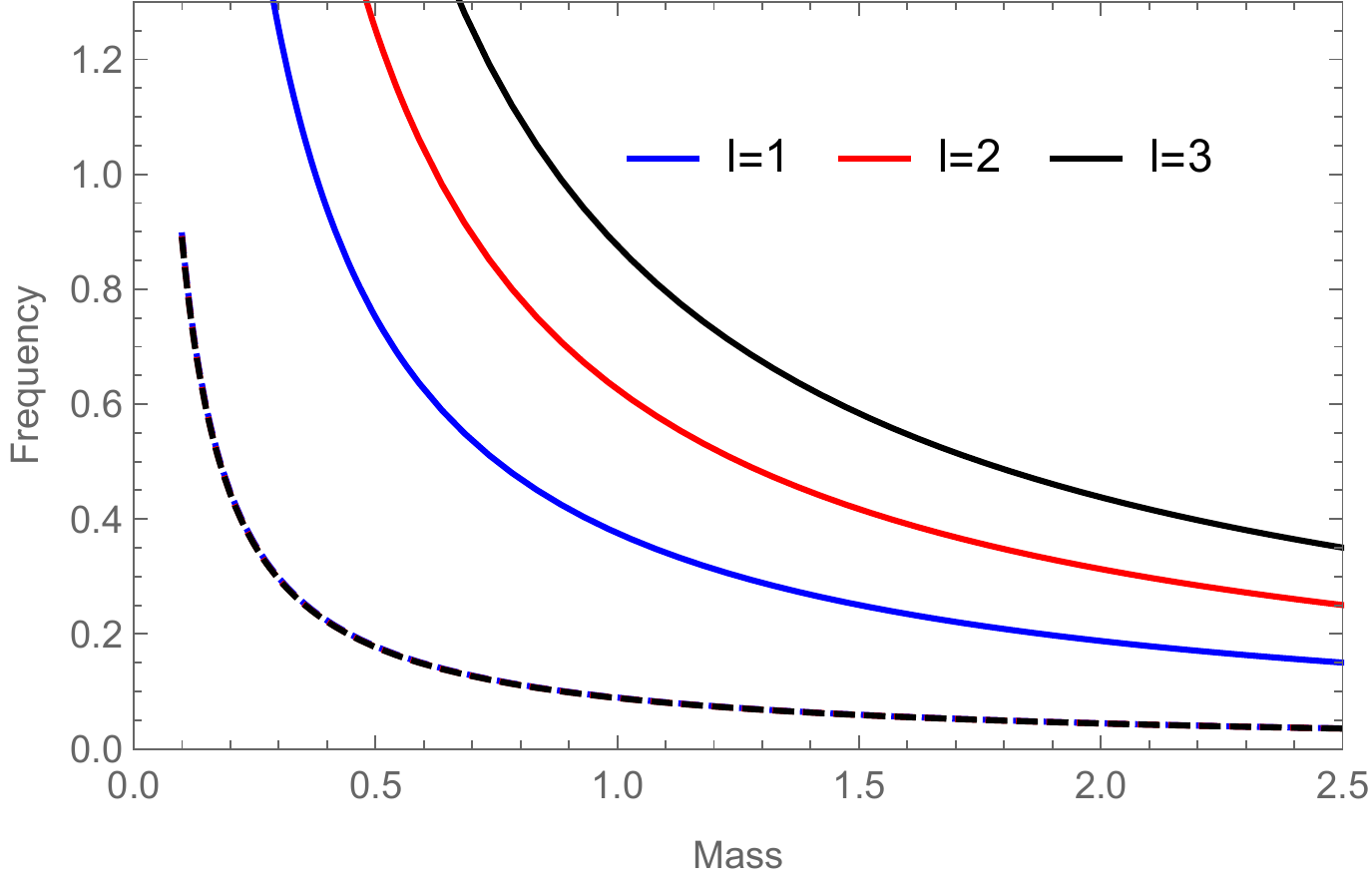} &  \includegraphics[width=\textwidth/2]{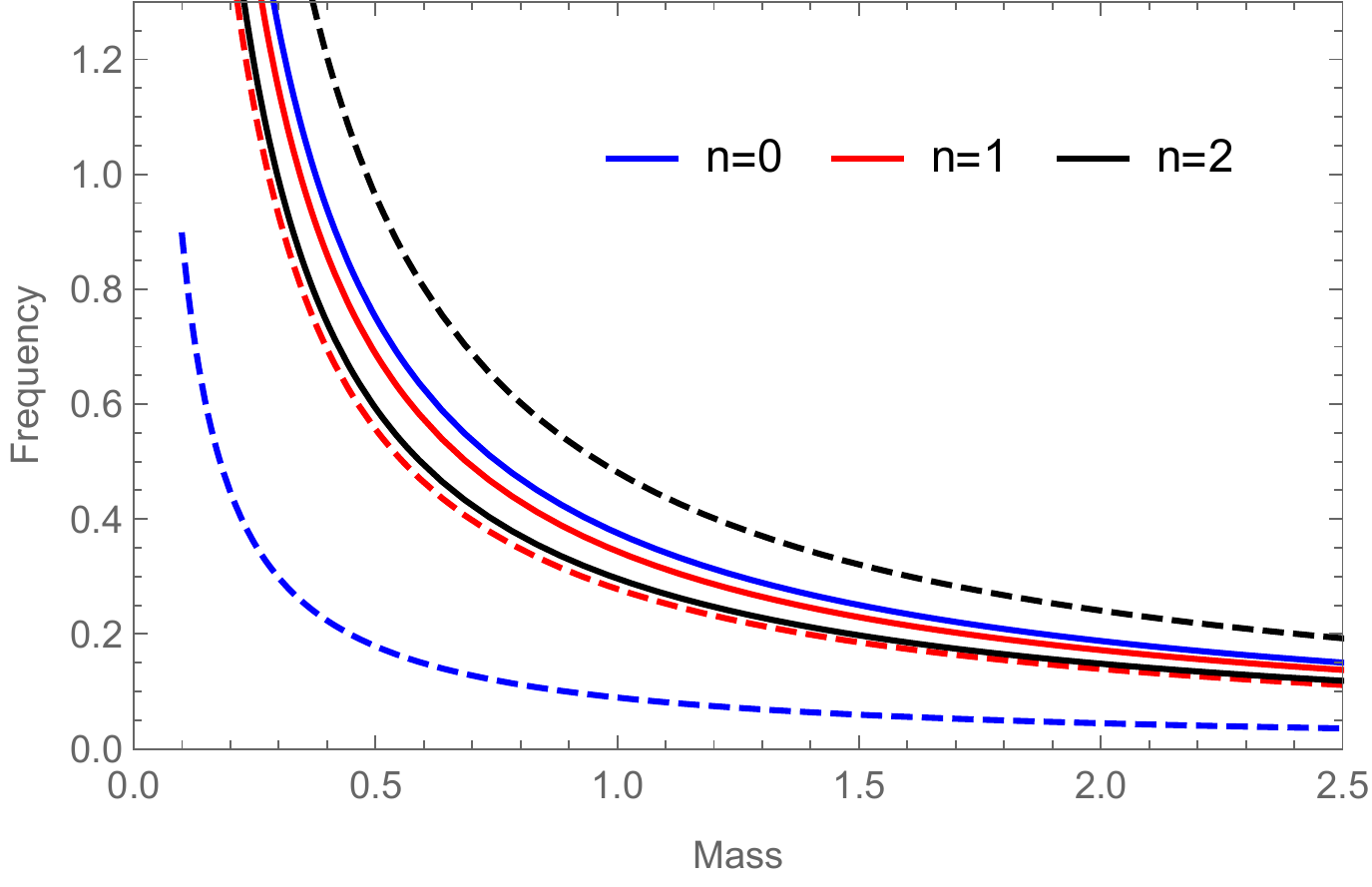}
  \end{tabular}
  \caption{\textit{Axial} perturbation of a $S=0$ field in Reissner-Norstr\"{o}m space-time with the extremal condition imposed, $Q = M$. The continuous lines represent the real frequencies and the dashed ones are the respective imaginary frequencies. The plot on the left maintains $n=0$ constant while $\ell$ varies through $1$ to $3$, while The plot on the right maintains $\ell = 1$ constant while $n$ varies through $0$ to $2$. The analysis of this case roughly follows the one for the Schwarzschild case.}
    \label{Plots: RNEs}
\end{figure*}

In figure \ref{Plots: RNEs}, we set the maximal condition, that is $Q=M$, for an \textit{axial} perturbation of a scalar field. We must emphasize that such condition can be analyzed through the WKB method because its effective potential satisfies all the conditions previously discussed in section \ref{Section 3}. We see that the frequencies behave well in the whole domain, with the mass/charge acting as a damping factor. It is curious to see that the extremal solution of a charged black hole behaves similar to the uncharged one, the Schwarzschild black hole; therefore the same analysis for the frequencies may be employed here.

As we have established for the Schwarzschild solution, the imaginary frequencies seem to be sensible only to the variation of $n$. This characteristic may be observed while keeping either the mass or the charge fixed, as displayed in both graphs.


In a nutshell, the Reissner-Nordstr{\"o}m solution is a stable black hole in which the mass damps its vibration. On the other hand, its charge slowly increases its vibration, while keeping the imaginary frequencies unaffected. Once again, the results obtained in our analysis reproduces what is known in the literature.

\subsection{Schwarzschild de-Sitter}
\label{section SDS}

We show the frequencies of the Schwarzschild de-Sitter space-time in figures \ref{Plots: SdSaScalar} through \ref{Plots: SdSpGrav}. We identified four main attributes in our analysis that are discussed below. 

\subsubsection*{i) Vanishing frequencies in the extremal case}
As we have reviewed in section \ref{Sds explanation}, for the exact value of $9\Lambda M^2 = 1$, the Schwarzschild de-Sitter space-time reaches its extremal case, where it has a unique horizon.

In figures \ref{Plots: SdSaScalar}, \ref{Plots: SdSaEM} and \ref{Plots: SdSpGrav} we see that the frequencies vanish when the extremal solution is reached. In this case, the extremal black hole does not emit any kind of radiation. Such result was expected, on account that it was previously proved that extremal solutions possess null Hawking temperature \cite{Cho:2008vr}, hence no further loss of energy is possible.

\subsubsection*{ii) Swap behaviour}
In the plots on the left side, where $n$ is constant, for figures \ref{Plots: SdSaEM} and \ref{Plots: SdSpGrav}, but most prominently in the former, we see a kind of a "mirror" effect on the frequencies. We notice that the real and imaginary frequencies swap their general behaviour upon reaching the extremal value.

Before this value, the real frequencies have three distinct curves, for each value of $\ell$, while the imaginary frequencies act as one curve; just like in the Schwarzschild case, $\ell$ hardly has an impact in their behaviour. After this point, the opposite happens and we see three different curves for the imaginary frequencies and one for the real ones, since these become sensible to only $n$.

Furthermore, the frequencies "swap sides" in respect to the $y$ direction. The real frequencies start with a higher value, when in comparison to the imaginary ones, but once the limiting case is reached the opposite occurs.

To the authors best knowledge this behaviour has not been previously discussed in the literature and deserves further investigation.

\subsubsection*{iii) Indication of a stable naked singularity with a positive mass}
Although not physical, the discussion whether naked singularities are stable, or not, has been of great theoretical interest. When the naked singularity occurs due to the negative mass, it was proven to be unstable for the cases of: Schwarschild \cite{Gleiser:2006yz}, Reissner-Nordstr{\"o}m \cite{Dotti:2006gc}\footnote{It was actually proven for positive mass, but the dependence goes with $M^2$, hence the same procedure applies for the negative case.}, Schwarzschild (anti) de-Sitter \cite{Cardoso:2006bv} and Kerr with positive mass \cite{Dotti:2008yr}.

\begin{figure*}[!ht]
  \begin{tabular}{cc}
  \includegraphics[width=\textwidth/2]{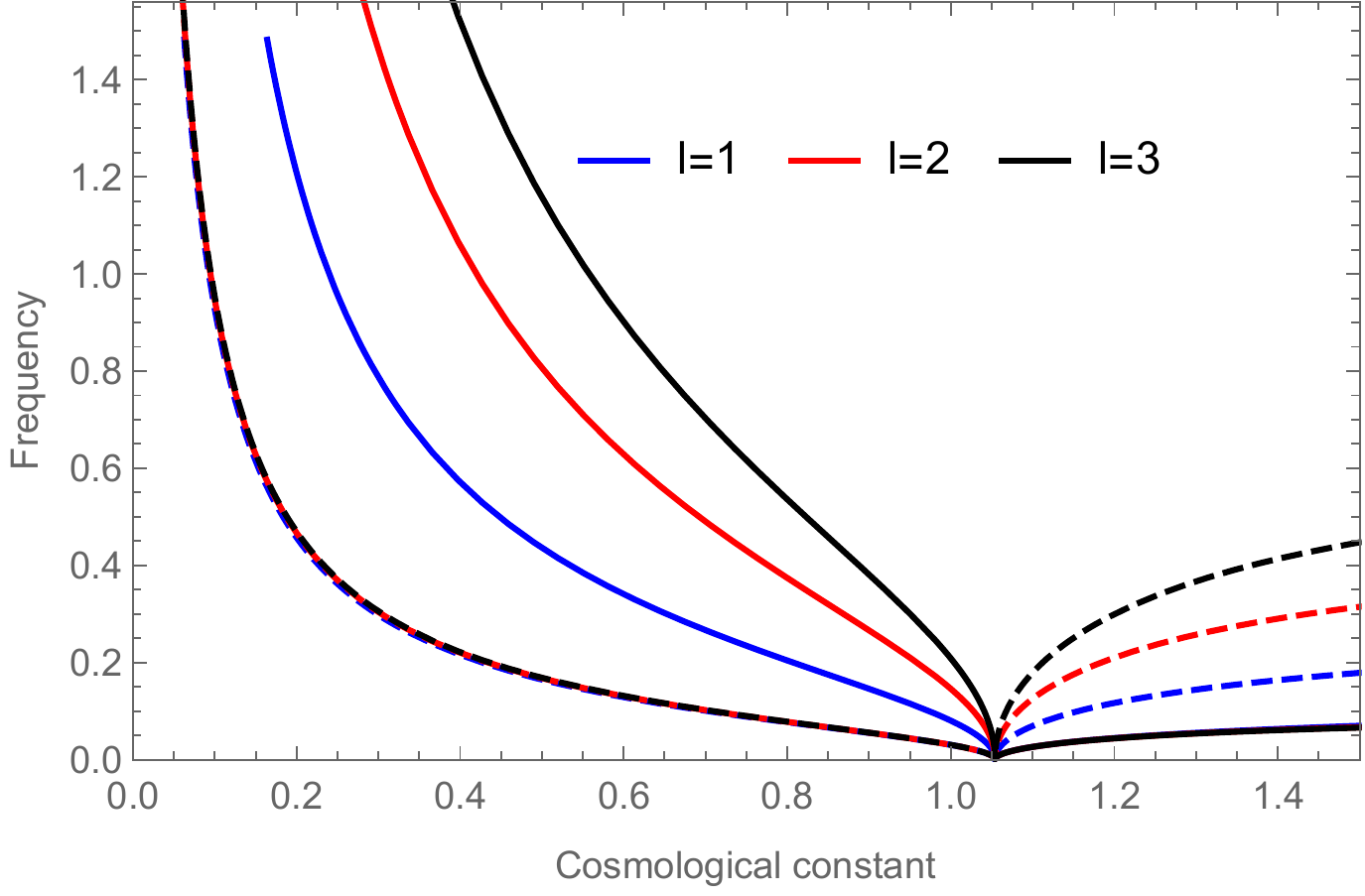} &  \includegraphics[width=\textwidth/2]{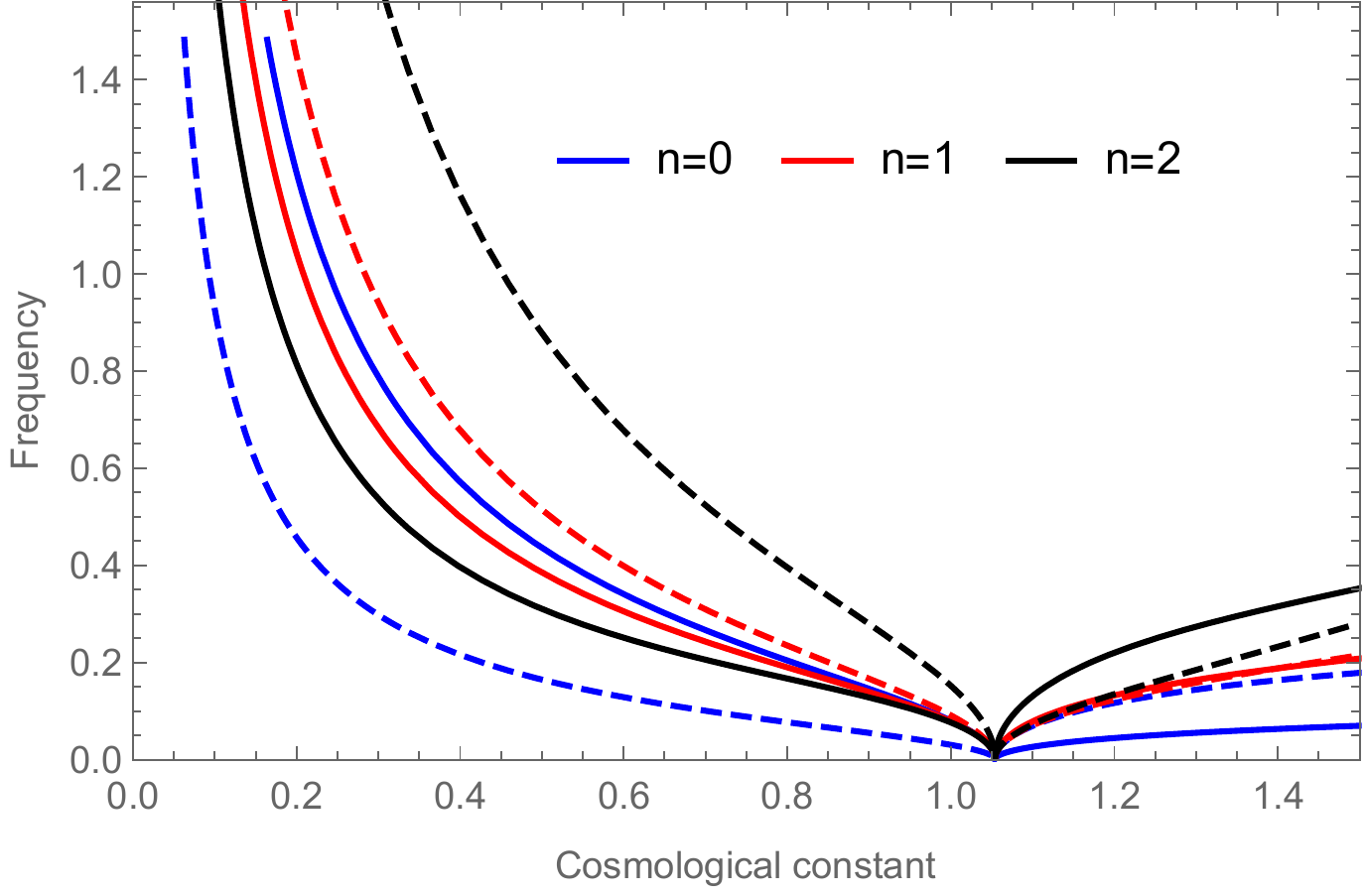}
  \end{tabular}
  \caption{\textit{Axial} perturbation of a $S=1$ field in Schwarzschild de-Sitter space-time for a fixed $M=1$. The continuous lines represent the real frequencies and the dashed ones are the respective imaginary frequencies. The plot on the left maintains $n=0$ constant while $\ell$ varies through $1$ to $3$. The plot on the right maintains $\ell = 1$ constant while $n$ varies through $0$ to $2$. We see that exactly the same behaviour as for the Schwarzschild cases follow before the frequencies vanish. On both plots it is possible to see the swap behaviour of the frequencies and the indication of a stable naked singularity, discussed in section \ref{section SDS}.}
  \label{Plots: SdSaEM}
\end{figure*}

\begin{figure*}[!ht]
  \begin{tabular}{cc}
  \includegraphics[width=\textwidth/2]{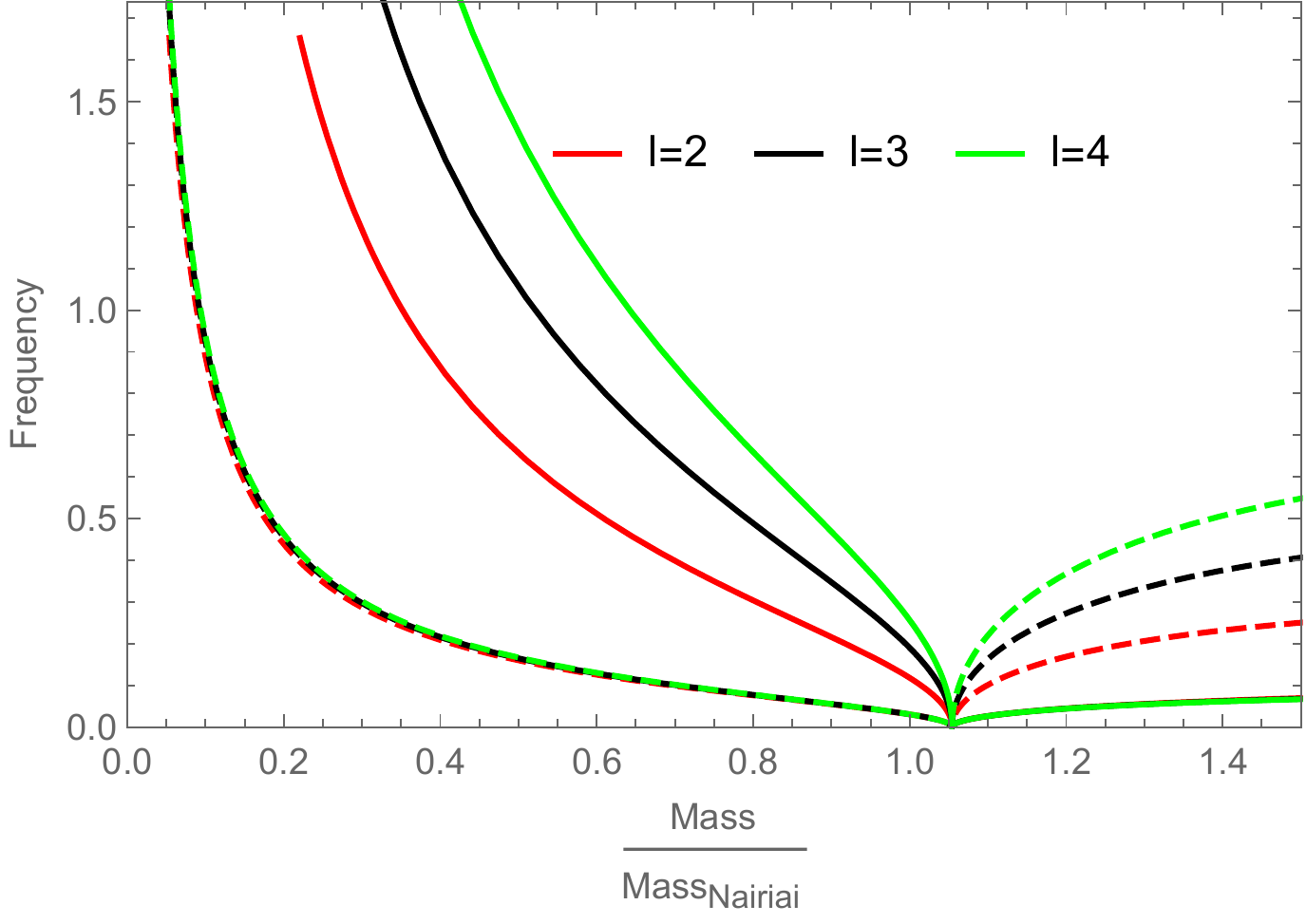} &  \includegraphics[width=\textwidth/2]{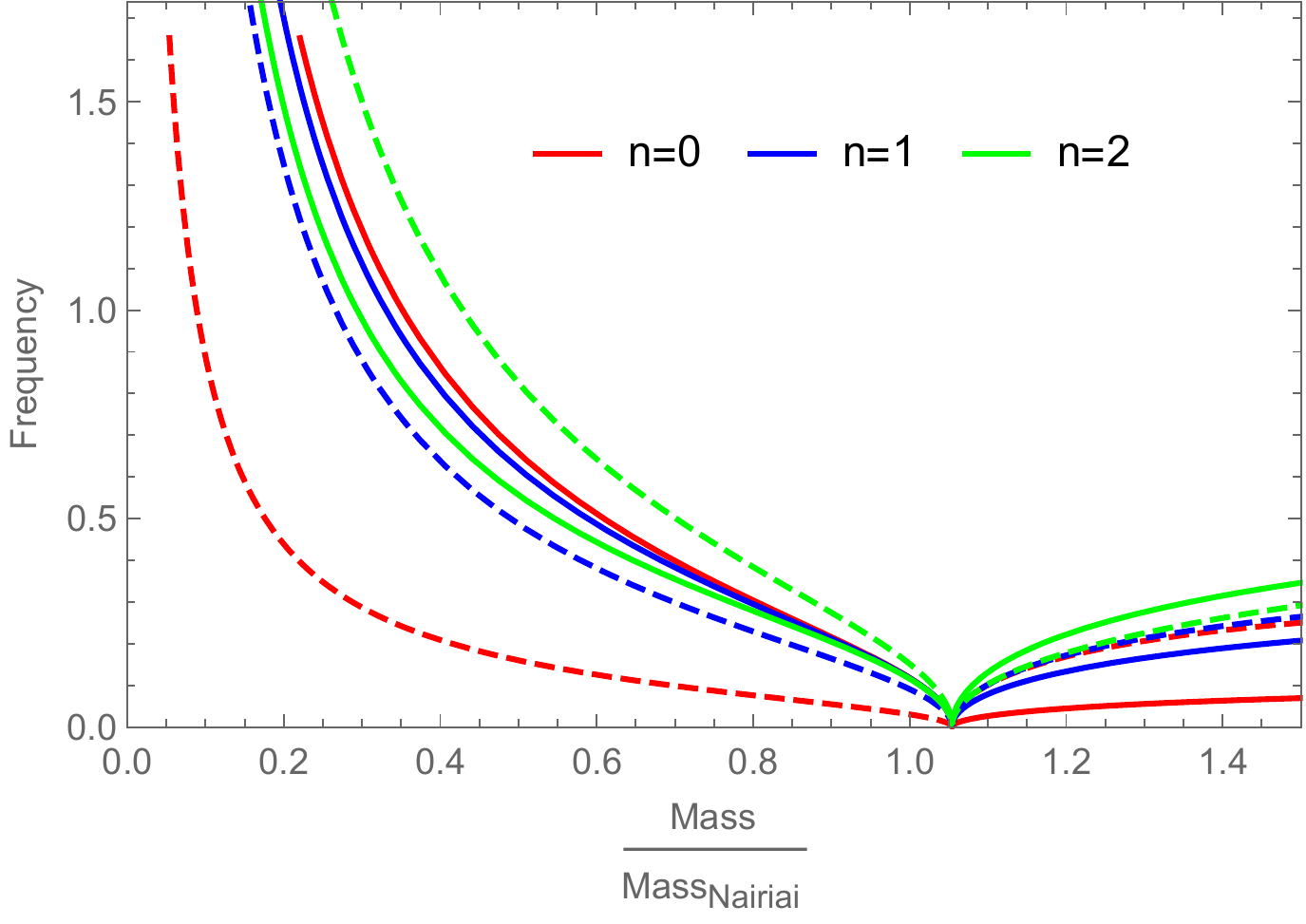}
  \end{tabular}
  \caption{\textit{Polar} perturbation of a $S=2$ field in Schwarzschild de-Sitter space-time for a fixed $\Lambda$, in terms of the mass divided by the Nairiai mass. The continuous lines represent the real frequencies and the dashed ones are the respective imaginary frequencies. The plot on the left maintains $n=0$ constant while $\ell$ varies through $2$ to $4$. The plot on the right keeps $\ell = 2$ constant while $n$ goes from $0$ to $2$. We see that exactly the same behaviour as for the Schwarzschild cases follow before the frequencies vanish. On both plots it is possible to see the swap behaviour of the frequencies and the indication of a stable naked singularity, discussed in section \ref{section SDS}.}
  \label{Plots: SdSpGrav}
\end{figure*}

Examining the figures \ref{Plots: SdSaEM} and \ref{Plots: SdSpGrav}, we see the existence of frequencies in the naked singularity domain, namely after the extremal case. Not only does this suggest that such a solution is stable, but it also becomes more stable as either $\Lambda$ or $M$ increases. The analysis we performed is not, by itself, a proof of the stability, but rather an indication that a more an in-depth investigation is necessary.

\subsubsection*{iv) Cut-off mass}
\begin{figure*}[!ht]
  \begin{tabular}{cc}
  \includegraphics[width=\textwidth/2]{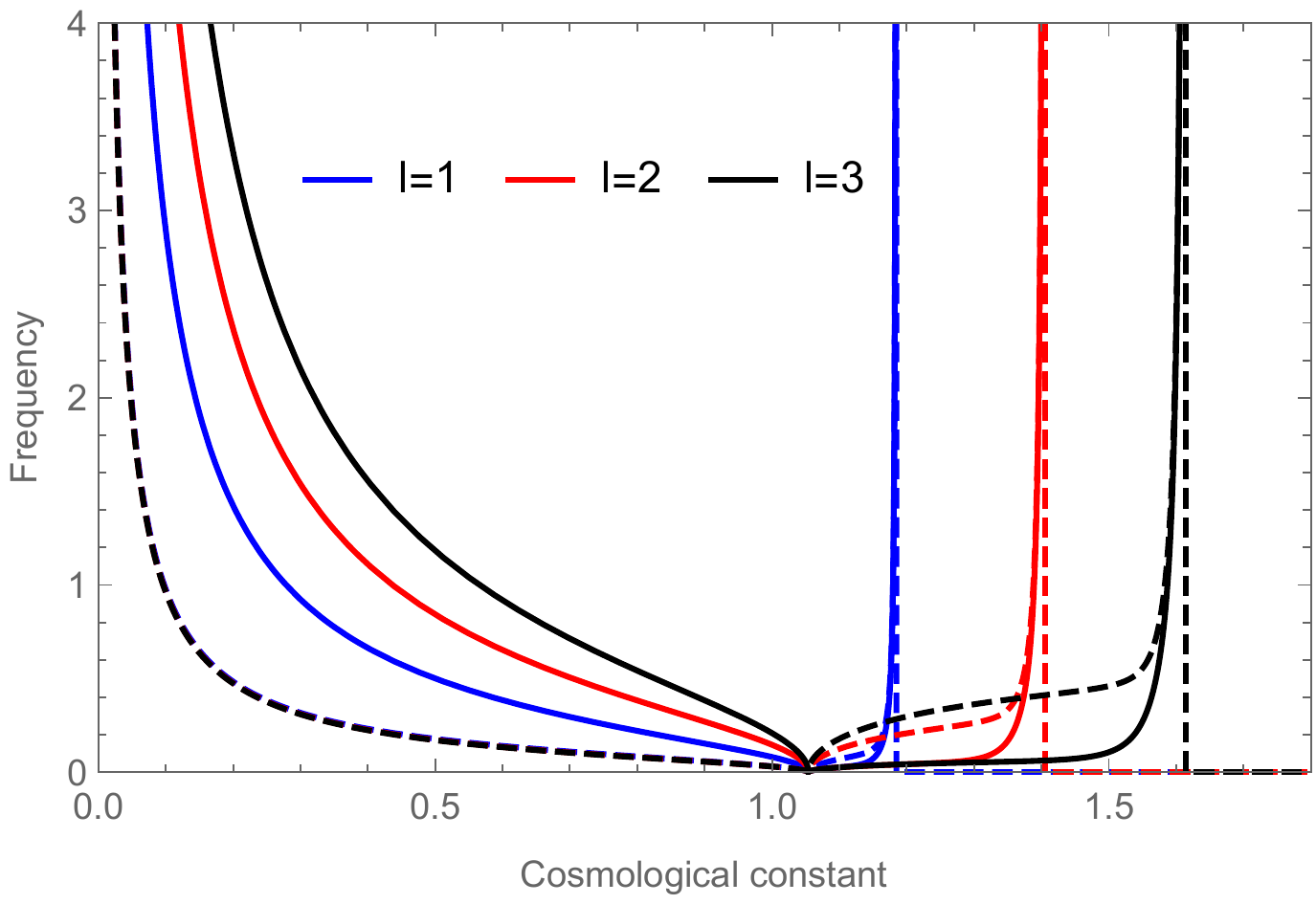} &  \includegraphics[width=\textwidth/2]{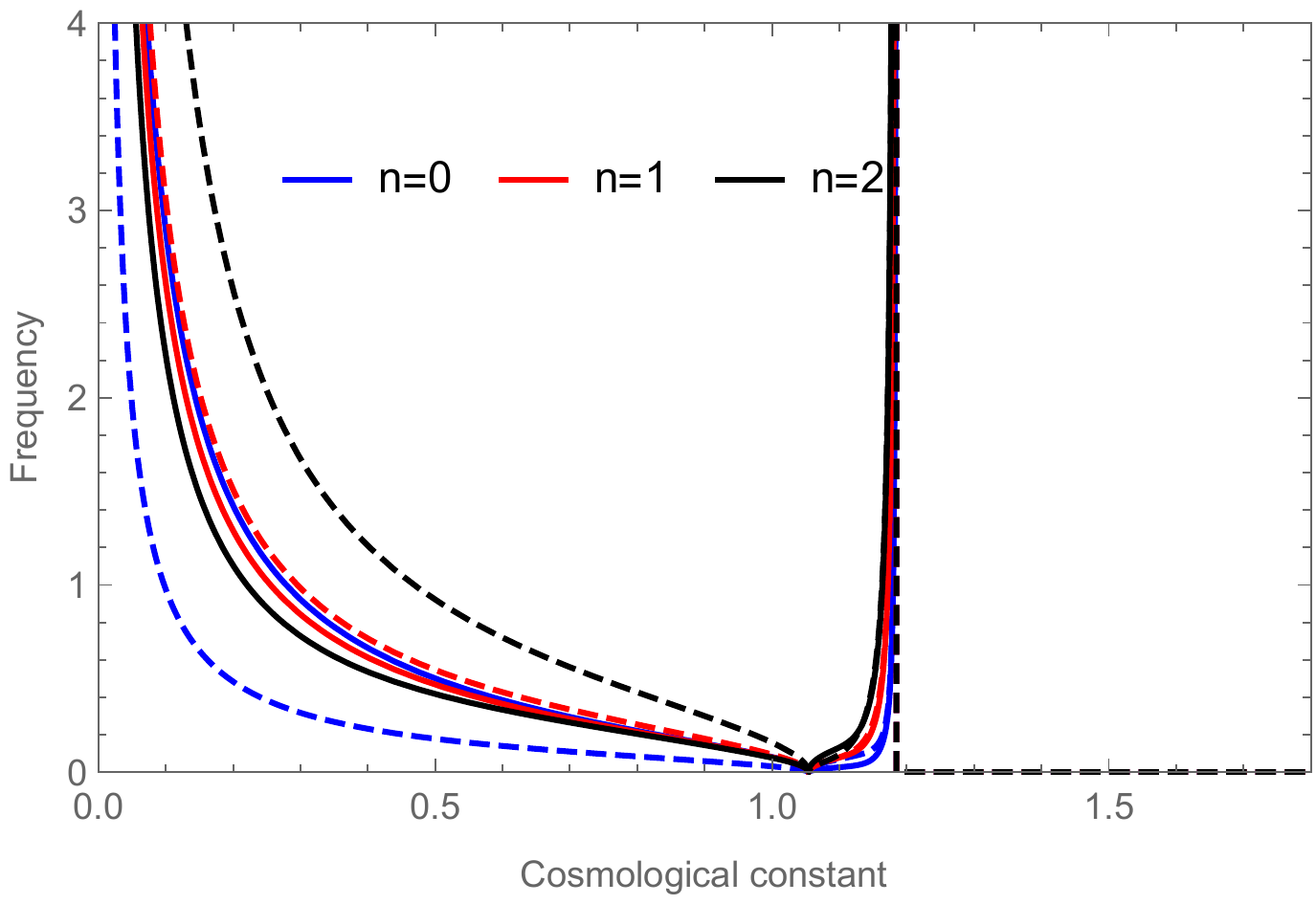}
  \end{tabular}
  \caption{\textit{Axial} perturbation of a $S=0$ field in Schwarzschild de-Sitter space-time for a fixed $M = 1$.The continuous lines represent the real frequencies and the dashed ones are the respective imaginary frequencies. The left plot maintains $n=0$ constant while $\ell$ varies through $1$ to $3$. The right plot maintains $\ell = 1$ constant while $n$ varies through $0$ to $2$. We see that exactly the same behaviour as for the Schwarzschild cases follow before the frequencies vanish. In the naked singularity part of both plots, it is possible to see that there exists a maximum value of the cosmological constant where the mode, for each $\ell$ but independent of $n$, diverges. The same behaviour occurs for the mass (not shown here), hence there are both a cut-off mass and a limit for $\Lambda$ for a scalar field in a naked singularity space-time.}
  \label{Plots: SdSaScalar}
\end{figure*}

The perturbation of a scalar-field is studied in figure \ref{Plots: SdSaScalar}. For this case, we see that for each mode, there is a value of $M$ beyond which no frequency exists.

On the plot on the left, we see that the cut-off mass does not depend on the value of the \textit{overtone} number, $n$, but solely on $\ell$. This is also true for the plot on the right.

It is important to stress that the exact same behaviour exists for a variable cosmological constant, thus it also has a cut-off $\Lambda$ value. \\ \\

Finally, we see that the black hole solution, for $0< 9 \Lambda M^2< 1$, is stable. The frequencies decrease when either $\Lambda$ or $M$ increase, and, consequently, both parameters act as damping factors.

\section{Conclusion}
\label{Summary}

In this work, we analysed the quasi-normal modes of the Schwarzschild, Reissner-Nordstr\"{o}m and Schwarzschild (anti) de-Sitter space-times, via the third-order WKB method, in the context of general relativity. Every black hole solution is stable under first-order perturbation, as it has been previously established using other methods. 

We checked that the Schwarzschild black hole is stable for any type of perturbation, with its mass acting as the damping factor.

We then verified that the charge of the Reissner-Nordstr\"{o}m black hole essentially keeps the imaginary frequencies constant, until just before reaching the extremal condition, as was first stated in \cite{Cardoso:2017soq}. We also note that the charge increases the vibration of the black hole, instead of damping it. The extremal case of the charged black hole was also examined, where we concluded that the frequencies behaves just like the Schwarzschild case, hence the same analysis applies.

The most important results we obtained were the four traits for the Schwarzschild de-Sitter solution: i) we showed that, as the black hole reaches its extremal status, both frequencies cease to exist, in agreement with the null Hawking temperature for these cases; ii) the frequencies swap their tendencies upon reaching $9 M^2 \Lambda$, where the real frequency acts like the imaginary one, and vice-versa; iii) we find an indication that the naked singularity may be stable for a positive mass; iv) lastly, we discovered the existence of a cut-off mass and cosmological constant, dependent on the \textit{overtone} number, in the naked singularity case of a scalar field perturbation. Although the first trait has already been thoroughly analysed in previous studies, the other three characteristics deserve a deeper investigation.

For every black hole solution studied here, we conclude that the imaginary frequencies, i.e. the damping factor of the oscillation, are insensible to the variation of $\ell$.

\vspace{1.3cm}

\noindent \textbf{Acknowledgements:} We are thankful to CAPES/Brazil, CPNq and FAPES for the financial support. The authors are grateful for the comments from Roman Konoplya. Edison C. Santos thanks Gustavo Dotti and Tassia Ferreira for useful discussions.

\clearpage

\bibliographystyle{apalike}
\balance
\bibliography{library}

\begin{thebibliography}{}

\bibitem[Bender and Orszag, 1978]{WKBbook}
Bender, C.~M. and Orszag, S.~A. (1978).
\newblock {\em Advanced Mathematical Methods for Scientists and Engineers I},
  volume~1.
\newblock Springer-Verlag New York, 1 edition.

\bibitem[Berti et~al., 2009]{Berti2009}
Berti, E., Cardoso, V., and Starinets, A.~O. (2009).
\newblock {Quasinormal modes of black holes and black branes}.
\newblock {\em Classical and Quantum Gravity}, 26(16).

\bibitem[Cardoso and Cavaglia, 2006]{Cardoso:2006bv}
Cardoso, V. and Cavaglia, M. (2006).
\newblock {Stability of naked singularities and algebraically special modes}.
\newblock {\em Phys. Rev.}, D74:024027.

\bibitem[Cardoso et~al., 2018]{Cardoso:2017soq}
Cardoso, V., Costa, J.~L., Destounis, K., Hintz, P., and Jansen, A. (2018).
\newblock {Quasinormal modes and Strong Cosmic Censorship}.
\newblock {\em Phys. Rev. Lett.}, 120(3):031103.

\bibitem[Cardoso and Lemos, 2001]{Cardoso:2001bb}
Cardoso, V. and Lemos, J. P.~S. (2001).
\newblock {Quasinormal modes of Schwarzschild anti-de Sitter black holes:
  Electromagnetic and gravitational perturbations}.
\newblock {\em Phys. Rev.}, D64:084017.

\bibitem[{Chandrasekhar}, 1983]{Chandrasekhar_book}
{Chandrasekhar}, S. (1983).
\newblock {\em {The mathematical theory of black holes}}.
\newblock The Oxford University Press.

\bibitem[Chandrasekhar, 1984]{Chandrasekhar1984}
Chandrasekhar, S. (1984).
\newblock {On Algebraically Special Perturbations of Black Holes}.
\newblock {\em Proceedings of the Royal Society A: Mathematical, Physical and
  Engineering Sciences}, 392(1802):1--13.

\bibitem[Chandrasekhar and Detweiler, 1975]{Chandrasekhar1975b}
Chandrasekhar, S. and Detweiler, S. (1975).
\newblock {The Quasi-Normal Modes of the Schwarzschild Black Hole}.
\newblock {\em Proceedings of the Royal Society A: Mathematical, Physical and
  Engineering Sciences}, 344(1639):441--452.

\bibitem[Cho et~al., 2010]{Cho:2008vr}
Cho, J.-H., Ko, Y., and Nam, S. (2010).
\newblock {The Entropy Function for the extremal Kerr-(anti-)de Sitter Black
  Holes}.
\newblock {\em Annals Phys.}, 325:1517--1536.

\bibitem[Dotti et~al., 2007]{Dotti:2006gc}
Dotti, G., Gleiser, R., and Pullin, J. (2007).
\newblock {Instability of charged and rotating naked singularities}.
\newblock {\em Phys. Lett.}, B644:289--293.

\bibitem[Dotti et~al., 2008]{Dotti:2008yr}
Dotti, G., Gleiser, R.~J., Ranea-Sandoval, I.~F., and Vucetich, H. (2008).
\newblock {Gravitational instabilities in Kerr space times}.
\newblock {\em Class. Quant. Grav.}, 25:245012.

\bibitem[Faraoni, 2018]{Faraoni:2018xwo}
Faraoni, V. (2018).
\newblock {Embedding black holes and other inhomogeneities in the universe in
  various theories of gravity: a short review}.
\newblock {\em Universe}, 4(10):109.

\bibitem[Fernando and Manning, 2017]{Fernando:2017qrd}
Fernando, S. and Manning, A. (2017).
\newblock {Electromagnetic perturbations of a de Sitter black hole in massive
  gravity}.
\newblock {\em Int. J. Mod. Phys.}, D26(09):1750100.

\bibitem[Gleiser and Dotti, 2006]{Gleiser:2006yz}
Gleiser, R.~J. and Dotti, G. (2006).
\newblock {Instability of the negative mass Schwarzschild naked singularity}.
\newblock {\em Class. Quant. Grav.}, 23:5063--5078.

\bibitem[Guven and N\'u\~nez, 1990]{PhysRevD.42.2577}
Guven, J. and N\'u\~nez, D. (1990).
\newblock Schwarzschild-de sitter space and its perturbations.
\newblock {\em Phys. Rev. D}, 42:2577--2584.

\bibitem[Hawking and Ellis, 1973]{Hawkingbook}
Hawking, S.~W. and Ellis, G. F.~R. (1973).
\newblock {\em {The Large Scale Structure of Space-Time}}.
\newblock Cambridge Monographs on Mathematical Physics. Cambridge University
  Press.

\bibitem[{Hidekazu}, 1951]{1951SRToh..35...46H}
{Hidekazu}, N. (1951).
\newblock {On a new cosmological solution of Einstein's field equations of
  gravitation}.
\newblock {\em Sci.~Rep.~Tohoku Univ.~Eighth Ser.}, 35.

\bibitem[Israel, 1967]{PhysRev.164.1776}
Israel, W. (1967).
\newblock Event horizons in static vacuum space-times.
\newblock {\em Phys. Rev.}, 164:1776--1779.

\bibitem[Iyer, 1987]{Iyer1987}
Iyer, S. (1987).
\newblock {Black Hole Normal Modes: a WKB Approach. II. Schwarzschild Black
  Holes}.
\newblock {\em Phys. Rev.}, D35(2004):3632.

\bibitem[Iyer and Will, 1987]{Iyer1987a}
Iyer, S. and Will, C.~M. (1987).
\newblock {Black-hole normal modes: A WKB approach. I. Foundations and
  application of a higher-order WKB analysis of potential-barrier scattering}.
\newblock {\em Physical Review D}, 35(12):3621--3631.

\bibitem[Kokkotas and Schmidt, 1999]{Kokkotas1999}
Kokkotas, K.~D. and Schmidt, B.~G. (1999).
\newblock {Quasi-normal modes of stars and black holes}.
\newblock {\em Living Reviews in Relativity}, 2.

\bibitem[Kokkotas and Schutz, 1988]{Kokkotas1988}
Kokkotas, K.~D. and Schutz, B.~F. (1988).
\newblock {Black-hole normal modes: A WKB approach. III. the
  reissner-nordstr{\"{o}}m black hole}.
\newblock {\em Physical Review D}, 37(12):3378--3387.

\bibitem[Konoplya, 2003]{Konoplya2003}
Konoplya, R.~A. (2003).
\newblock {Quasinormal behavior of the D-dimensional Schwarzschild black hole
  and the higher order WKB approach}.
\newblock {\em Physical Review D}, 68(2):1--8.

\bibitem[Konoplya and Zhidenko, 2011]{Konoplya2011}
Konoplya, R.~A. and Zhidenko, A. (2011).
\newblock {Quasinormal modes of black holes: From astrophysics to string
  theory}.
\newblock {\em Reviews of Modern Physics}, 83(3):793--836.

\bibitem[Kottler, 1918]{Kottler1918}
Kottler, F. (1918).
\newblock Über die physikalischen grundlagen der einsteinschen
  gravitationstheorie.
\newblock {\em Annalen der Physik}, 361(14):401--462.

\bibitem[Landau and Lifshitz, 1981]{Landau_QM}
Landau, L.~D. and Lifshitz, E.~M. (1981).
\newblock {\em Quantum Mechanics}, volume~3.
\newblock Butterworth-Heinemann, 3 edition.

\bibitem[Leaver, 1985]{Leaver1985}
Leaver, E.~W. (1985).
\newblock {An Analytic Representation for the Quasi-Normal Modes of Kerr Black
  Holes}.
\newblock {\em Proceedings of the Royal Society A: Mathematical, Physical and
  Engineering Sciences}, 402(1823):285--298.

\bibitem[Maldacena, 1999]{Maldacena:1997re}
Maldacena, J.~M. (1999).
\newblock {The Large N limit of superconformal field theories and
  supergravity}.
\newblock {\em Int. J. Theor. Phys.}, 38:1113--1133.
\newblock [Adv. Theor. Math. Phys.2,231(1998)].

\bibitem[{Mashhoon}, 1983]{1983mgm..conf..599M}
{Mashhoon}, B. (1983).
\newblock {Quasi-normal modes of a black hole.}
\newblock In {Ning}, H., editor, {\em Third Marcel Grossmann Meeting on General
  Relativity}, pages 599--608.

\bibitem[Nikolaos, 2005]{Nikolaos2005}
Nikolaos, F. (2005).
\newblock {Quantizing Gravity : Insights from Quasinormal Modes of Black
  Holes}.
\newblock (June):79.

\bibitem[Nollert, 1999]{Nollert1999}
Nollert, H.-p. (1999).
\newblock {Quasinormal modes : the characteristic ` sound ' of black holes and
  neutron stars Quasinormal modes : the characteristic ‘ sound ’ of black
  holes and neutron stars}.
\newblock 159.

\bibitem[Penrose, 1969]{Penrose:1969pc}
Penrose, R. (1969).
\newblock {Gravitational collapse: The role of general relativity}.
\newblock {\em Riv. Nuovo Cim.}, 1:252--276.
\newblock [Gen. Rel. Grav.34,1141(2002)].

\bibitem[Regge and Wheeler, 1957]{Regge1957}
Regge, T. and Wheeler, J.~A. (1957).
\newblock {Stability of a schwarzschild singularity}.
\newblock {\em Physical Review}, 108(4):1063--1069.

\bibitem[Schutz and Will, 1985]{Schutz}
Schutz, B.~F. and Will, C.~M. (1985).
\newblock {Black hole normal modes - A semianalytic approach}.
\newblock {\em The Astrophysical Journal}, 291:L33.

\bibitem[Seidel and Iyer, 1990]{Seidel1990}
Seidel, E. and Iyer, S. (1990).
\newblock {Black-hole normal modes: A WKB approach. IV. Kerr black holes}.
\newblock {\em Physical Review D}, 41(2):374--382.

\bibitem[Teukolsky, 1973]{Teukolsky:1973ha}
Teukolsky, S.~A. (1973).
\newblock {Perturbations of a rotating black hole. 1. Fundamental equations for
  gravitational electromagnetic and neutrino field perturbations}.
\newblock {\em Astrophys. J.}, 185:635--647.

\bibitem[Vishveshwara, 1970]{Vishveshwara1970}
Vishveshwara, C.~V. (1970).
\newblock {Stability of the Schwarzschild metric}.
\newblock {\em Physical Review D}, 1(10):2870--2879.

\bibitem[Zerilli, 1970]{Zerilli1970}
Zerilli, F.~J. (1970).
\newblock {Gravitational field of a particle falling in a schwarzschild
  geometry analyzed in tensor harmonics}.
\newblock {\em Physical Review D}, 2(10):2141--2160.

\bibitem[Zhidenko, 2004]{Zhidenko2003}
Zhidenko, A. (2004).
\newblock {Quasi-normal modes of Schwarzschild-de Sitter black holes}.
\newblock {\em Classical and Quantum Gravity}, 21(1):273--280.

\end{thebibliography}

\end{document}